\documentclass[preprint,12pt]{elsarticle}
\journal{Annals of Physics}
\usepackage{graphicx}

\usepackage{epstopdf}

\begin{document}


\title{Neutron star under homotopy perturbation method}

\author{Abdul Aziz$^a$, Saibal Ray$^{b}\footnote{$^*$Corresponding author.\\
{\it E-mail addresses:} aziz.rs2016@physics.iiests.ac.in (AA), saibal@associates.iucaa.in (SR), rahaman@associates.iucaa.in (FR), dean.fa@iiests.ac.in (BKG).}$, Farook Rahaman$^c$, B.K. Guha$^a$}

\address{$^a$Department of Physics, Indian Institute of Engineering Science and Technology, Shibpur, Howrah, West Bengal, 711103, India\\
$^b$Department of Physics, Government College of Engineering and Ceramic Technology, Kolkata 700010, West Bengal, India \& Department of Natural Sciences, Maulana Abul Kalam Azad University of Technology, Haringhata 741249, West Bengal, India\\
$^c$Department of Mathematics, Jadavpur University, Kolkata 700032, West Bengal, India}
\date{Received: date / Accepted: date}

\maketitle

\begin{abstract}
We obtain a mass function solving the Tolman-Oppenheimer-Volkoff (TOV) equation for isotropic and spherically symmetric system via homotopy perturbation method (HPM). Using the mass function we construct a stellar model which can be determined from the equation of state (EOS) parameter ($\omega$) and a model parameter ($n$). With the help of Einstein field equations we develop three solutions which can describe different properties and the core-crust structure of neutron star (NS). Solution I is valid for NS having the inner and outer radius near the surface of the star. The star is physical up to the inner radius whereas negative density occurs and the energy conditions are violated in the upper region from the inner to outer radius. Solution II represents NS with high gravitational redshift as well as compactification factor. All the features of NS can be given by solution III which involves only the EOS parameter. Our model predicts maximum mass for a NS with the central density $5.5\times10^{15}~g/cm^{3}$ and surface redshift 0.69 is to be $2.01~M_\odot$ for the EOS parameter $\omega=0.73$. The predicted range for the surface redshift is $0.57<Z_s<1.95$ for the allowed ranges $8.4<n<10.9$ and $1/3<\omega<1$ in the presented NS model. 
\end{abstract}


\section{Introduction}
One of the best ways to test Einstein's general relativity in the regime of strong gravitational field is to study the properties of compact stars, especially to investigate neutron star (NS) as a compact star candidate. The detection of Gravitational wave ($GW~170817$) and gamma ray burst ($GRB~170817A$) from a binary NS merger~\cite{Abbott2017a} by the detectors of LIGO and Virgo is very much promising to explore many more facts in astrophysics as well as in cosmology~\cite{Abbott2017b}. The phenomena of merger of NSs depend on not only the EOS of NS but also the mass-radius relation~\cite{Lattimer2000}. So, the gravitational wave and electromagnetic wave coming from NSs carry the necessary information regarding the structural properties of the star.

According to many theoretical models and experimental observations, the structure of NS  consists of core and crust region. The density in the core region of NS can be from as low as $3 \times 10^{14}~g/cm^{3}$ to as high as $3 \times 10^{15}~g/cm^{3}$. The core density is comparable and sometimes greater than the normal nuclear density which is about $\rho_{0}=2.7 \times 10^{14}~g/cm^{3}$. At the core-crust boundary of the NS the density and pressure are up to $ \rho_{cc} \approx(0.3-0.5)\rho_{0}$  and  $p_{cc} \approx 10^{33}~dyn/cm^{2}$ respectively~\cite{Haensel2007}. The equation of state for a star is soft for compressible fluids  and stiff for incompressible fluids which are basically governed by the functional relationship between the pressure and density of star. The range for mass of NS is $ M\approx 1- 3~M_\odot$~\cite{Miller2015} (for review). Theoretically upper limit for maximum mass of NS is about $3.2~M_\odot$~\cite{Rhoades1974}.  For sub-nuclear density region ($\rho< \rho_{0}$), only nucleons are present in matter interacting with nuclear forces. This region contributes only 1-3 \% in mass for a NS of canonical mass $M=1.4~M_\odot$ and contributes less than one percent for massive NS of $M_{max}> 2~M_\odot$. The supranuclear density region ($\rho> \rho_{0}$) consists of dense matter undergoing strong interactions and contributes maximum in mass formation of NS. 

However, it is very difficult to predict exact equation of state because many-body theories of strong interactions for dense matter is yet not successfully developed in nuclear physics. But some theories are proposed, such as Brueckner-Bethe-Goldstone theory, according to which the maximum mass for neutron star of nucleon-hyperon core above supranuclear density with very high central density is up to $1.5~M_\odot$ \cite{Schulze2011}. This upper limit of mass cannot be matched with recent discovery of massive pulsars \cite{Barziv2001,Freire2008,Demorest2010,Kerkwijk2011,Romani2012,Antoniadis2013}. To overcome this conflict, it is prescribed that in addition of much stiffer EOS either there will be strong repulsive forces between hyperons via exchange of $\phi$-meson \cite{Dexheimer2008,Bednarek2012,Weissenborn2012a,Bonanno2012} or there will be strong SU(6) symmetry breaking \cite{Weissenborn2012b}. Based on this prescription there are models using relativistic mean theory which predicts $M_{max}> 2~M_\odot$ for nucleon-hyperon core of NS. 

Again the maximum mass of NS of nucleon core at $\rho_{0}<\rho \leq 2\rho_{0}$ can be greater than two solar mass \cite{Chamel2013}. It is to be noted that Quantum Chromodynamics allows deconfinement of quarks in baryon core at certain density. At this point the core matter become plasma of quarks which strongly interacts with exchange of gluons. The NSs whose core consists of baryons (confined state of quarks)  and quark-gluon plasma is called hybrid star which could be of  $M_{max}> 2~M_\odot$~\cite{Blaschke2010,Weissenborn2011,Zdunik2013,Colucci2013}. In these stars phase transition occur from baryon to quark matter. The quark matter is very stiff so that sound speed is about $(0.8-0.9)~c$.  For detailed study of neutron structure one may consult the following literature~\cite{Lattimer2004,Lattimer2007,Haensel2016,Eksi2016}.

It can be noted that the discovery of massive NSs imposes constraints through the equation of state. For stiff equation of state, where the speed of sound will be very close to speed of light, NS could be massive as much as 2~$M_\odot$~\cite{Steiner2010}. It is predicted that NS may born massive following different evolutionary process or extra mass may be accreted~\cite{Frank2002} from companion in X-ray binary system. The high nuclear density at the core region  also plays crucial role in determination of the mass and radius of NS. With the help of Shaprio delay mass measurements Demorest et al. \cite{Demorest2010} have shown that soft EOS such as  hyperon or kaon condensation at nuclear saturation densities  is not applicable for millisecond pulsar $PSR~J1614−2230$. {\"O}zel et al.~\cite{Ozel2016} using recent measurements with uncertainty predict that the radii range for NS is 9.9 - 11.2~km. The systematic errors in measurement of NS radii from thermonuclear bursts are about 3 - 8 \%~\cite{Guver2012}. Gravitational wave detection, neutrino emission and measurement of moment of inertia are expected to help more to explore the NS structure. 

We note that there is still uncertainty in measurements of maximum mass and radius of NS. Therefore our motivation here is to study the properties of NS and to find all possible non-singular solutions for isotropic stellar structure via homotopy perturbation method. Basically in the present study we try to solve the TOV equation and obtain a mass function. In nonlinear science some nonlinear equations are often not solvable analytically where an approximate method is applicable. We get approximate analytic solutions for the TOV using homotopy perturbation method which has first proposed by He~\cite{He1997,He1999,He2000a,He2000b,He2004,He2005,He2006,He2010}. This method is a powerful and simple technique which can reduce nonlinear problems to simpler one. Generally the solution is in series form but it converges very rapidly making the approximate solutions more physical and very interesting. The homotopy perturbation method has been successfully used in the field of astrophysics and cosmology~\cite{Shchigolev2013,Shchigolev2015,Shchigolev2016,Shchigolev2017}.

Therefore, under the above background this paper is devoted entirely to an analysis of the Tolman-Oppenheimer-Volkoff (TOV) equation with the linear equation of state of the form $p = \omega \rho$, where $p$ is the pressure, $\rho$ is the energy density, and $\omega$ is a constant. We have constructed and discussed here a set of approximate regular solutions. The plan of the work is as follows: In Section~2 we derive a nonlinear differential equation for mass from TOV equation using Einstein's field equations for spherically symmetric line element and isotropic prefect fluid. We give brief idea of homotopy perturbation method and derive a mass function in Section~3. With the help of this mass function we develop star model and study different features, such as density, pressure, radius, total mass, compactness, redshift of star, time-time component of metric, core-crust structure in Section~4. Important results of the solutions of the model are discussed in Section~5. Then in Section~6, the solutions are put to test for physical validity. A detailed discussion on Mass-Radius relation is done in Section~7. At last we made some important remarks in conclusion part of Section~8.

\section{Spherically symmetric spacetime and TOV equation}  
Let us consider the  spherically symmetric spacetime with following metric tensor
\begin{equation}
ds^{2}=-g_{tt}(r)dt^{2}+\left(1-\frac{2m(r)}{r}\right)^{-1}dr^{2}+r^{2}(d\theta^{2}+sin^{2}\theta d\phi^{2}), \label{eq1}
\end{equation}
along with the energy momentum tensor for  perfect fluid (taking isotropic condition which means the pressure of the system in the radial and tangential direction will be same)
\begin{equation}
T_\nu^\mu=  ( \rho + p)u^{\mu}u_{\nu} + p g^{\mu}_{\nu}, \label{eq2}
\end{equation}
with $u^{\mu}u_{\mu} = 1$, where $u^{\mu}$ is the 4-fluid velocity.

We take linear equation of state for the fluid as 
\begin{equation}
p = \omega \rho, \label{eq3}
\end{equation}
which shows that the pressure is directly proportional to the density and the proportionality constant $\omega$ is called EOS parameter.

The Einstein field equations are given by 
\begin{equation}
\frac{2m^{\prime}}{r^{2}}=8\pi\rho, \label{eq4}
\end{equation}

\begin{equation}
\frac{2m}{r^{3}} -
\left(1-\frac{2m}{r}\right)\frac{g_{tt}^{\prime}}{g_{tt}}\frac{1}{r}
= -  8\pi p, \label{eq5}
\end{equation}

\begin{eqnarray}
-\left(1-\frac{2m}{r}\right)\left[\frac{1}{2}\frac{g_{tt}^{\prime\prime}}{g_{tt}}
-\frac{1}{4}\left(\frac{g_{tt}^{\prime}}{g_{tt}}\right)^{2} +
\frac{1}{2r}\frac{g_{tt}^{\prime}}{g_{tt}}\right]   -\left (\frac{m}{r^{2}}-\frac{m^{\prime}}{r}\right)
\left[\frac{1}{r} +
\frac{1}{2}\frac{g_{tt}^{\prime}}{g_{tt}}\right] = - 8\pi p,\label{eq6}
\end{eqnarray}
where the gravitational constant ($G$) and the speed of light ($c$) are taken to be unity. 

From the energy conservation relation, which implies covariant derivative of energy momentum tensor will be zero, i.e., $\nabla_{\nu} T^{\mu \nu}=0$, we get the following relation
\begin{equation}
p^{\prime}= - \frac{(\rho + p )g_{tt}^{\prime}}{2 g_{tt}}. \label{eq7}
\end{equation}
Using Eq. (\ref{eq5}) and Eq. (\ref{eq7}), we finally get the  TOV equation \cite{Tolman1939,Oppenheimer1939}
\begin{equation}
p^{\prime}= - \frac{(\rho + p)(m + 4\pi r^{3}p)}{r(r-2m)}. \label{eq8}
\end{equation}
It is to be noted that from a purely theoretical point of view, Chandrasekhar~\cite{Chandrashekhar1972} did analysis of this TOV equation which contains important elements: a derivation of an elegant, dimensionless form (Eqs. (16) and (17), p. 187) or an analysis of the asymptotic behaviour (p. 188). However, the analysis here is much different than that of Chandrasekhar and also the field of applicability differs widely as can be seen later on.

The above equation can be written in the following suitable form 
\begin{equation}
-\frac{M_G\left(\rho+p\right)}{r^2}e^{\frac{\lambda-\nu}{2}}-\frac{dp}{dr}=0, \label{eq9}
\end{equation}
where $M_G=M_G(r)$ is the Tolman-Whittakar mass given by
\begin{equation}
M_G(r)=\frac{1}{2}r^2e^{\frac{\nu-\lambda}{2}}\nu^{\prime}.\label{eq10}
\end{equation}
From Eq. (\ref{eq9})
\begin{equation}
 F_g+ F_h=0, \label{eq11}
\end{equation}
where
\begin{equation}
F_{g}=-(\omega+1)\rho\frac{g_{tt}^{\prime}}{2g_{tt}},\label{eq12}
\end{equation}

\begin{equation}
F_{h}=-\omega\frac{d\rho}{dr}.\label{eq13}
\end{equation}
Here Eq. (\ref{eq11}) represents the hydrostatic equilibrium of the stellar configuration. The stability of star
is maintained as the gravitational force ($F_g$) is balanced by the hydrostatic force ($F_h$).

Now using  Eqs. (\ref{eq3}) and (\ref{eq4}) in Eq. (\ref{eq8}), we get 
\begin{equation}
m^{\prime} - \frac{1}{2}m^{\prime\prime}r  + m^{\prime\prime}m -\frac{(5\omega+1)}{2\omega} \frac{mm^{\prime}}{r} -
\frac{(\omega+1)}{2}(m^{\prime})^{2}  = 0.\label{eq14}
\end{equation}
To determine the mass function we need to solve the above non linear differential 
equation for mass. For this we use homotopy perturbation method (HPM).

\section{Mass function via Homotopy Perturbation Method} 
We now discuss the HPM very briefly for which we consider the following nonlinear equation
\begin{equation}
A(u) - f(r)=0,\label{eq15}
\end{equation}
with the boundary condition $B\left(u,\frac{\partial u}{\partial n}\right)=0$. Here $A$ and $B$ are differential operator and  boundary operator respectively, $f(r)$ is a known analytical function and $\frac{\partial}{\partial n}$ is directional derivative. The basic idea is to deform the linear problem into nonlinear problem by building a suitable homotopic relation with introduction of an embedding parameter $\epsilon$. The first step is to divide the operator $A$ into linear part $L$ and nonlinear part $N$. So Eq. (\ref{eq15}) can be written
\begin{equation}
L(u) + N(u) - f(r)= 0. \label{eq16}
\end{equation}
Then  a homotopy structure should be put in form 
\begin{equation}
H(u,\epsilon)= L(u) -L(u_{0})+ \epsilon L(u_{0})+\epsilon[N(u)- f(r)], \label{eq17}
\end{equation}
where $u_{0}$ is an initial approximation. It is clear that if $\epsilon$ changes from 0 to 1, $H(u,0)=L(u)- L(u_{0})$ is continuously transformed into $H(u,1)=A(u) - f(r)$. The functions $H(u,0)$ and $H(u,1)$ are called homotopic functions.

We utilize the HPM method to solve  Eq. (\ref{eq14}). We construct homotopy relation such that 
 
\begin{equation}
m^{\prime} - \frac{1}{2}m^{\prime\prime}r  + \epsilon\left[m^{\prime\prime}m -\omega_{1} \frac{mm^{\prime}}{r} -
\omega_{2}(m^{\prime})^{2}\right]  = 0 ,\label{eq18}
\end{equation}
with
\[
\left(\frac{5}{2} + \frac{1}{2\omega}\right) =\omega_{1}~~and~~\left(\frac{1}{2} +\frac{\omega}{2}\right) = \omega_{2}.
 \]
We assume the mass solution as 
\begin{equation}
m = m_{0}+\epsilon m_{1}+\epsilon^2m_{2}...\label{eq19}
\end{equation}
Substituting Eq. (\ref{eq19}) into Eq. (\ref{eq18}) and equating the coefficients of $\epsilon$ to zero we get the following expressions:
 
$\epsilon^{0} :$ 
\begin{equation}
 m_{0}^{\prime} - \frac{1}{2}m_{0}^{\prime\prime}r =0, \label{eq20}
\end{equation}

$\epsilon^{1} :$ 
\begin{equation}
 m_{1}^{\prime} - \frac{1}{2}m_{1}^{\prime\prime}r + m_{0}^{\prime\prime}m_{0} -\omega_{1} \frac{m_{0}m_{0}^{\prime}}{r} -
\omega_{2}(m_{0}^{\prime})^{2} =0, \label{eq21} 
\end{equation}

$\epsilon^{2} :$ 
\begin{eqnarray}
 m_{2}^{\prime} - \frac{1}{2}m_{2}^{\prime\prime}r + m_{0}^{\prime\prime}m_{1}+ m_{1}^{\prime\prime}m_{0} \nonumber
 \\
  -\omega_{1} \frac{(m_{0}m_{1}^{\prime}+m_{1}m_{0}^{\prime})}{r} -
2\omega_{2}m_{0}^{\prime} m_{1}^{\prime} =0.\label{eq22}
\end{eqnarray}
and so on.

Solving Eqs. (\ref{eq20}), (\ref{eq21}) and (\ref{eq22}) we get 
\begin{equation}
m_{0}= C_{1} + C_{2}r^{3},\label{eq23}
\end{equation}

\begin{equation}
m_{1}= C_{2}^{2}\omega_{3} r^{5} +\frac{1}{3} C_{3}r^{3}+C_{4}, \label{eq24}
\end{equation}
with
\[
-\frac{3}{5} \left(2+\frac{3}{2}\omega+\frac{1}{2\omega}\right)=\omega_{3},
 \]

\begin{equation}
m_{2}= C_{2}^{3}\omega_{5} r^{7} + C_{2} C_{3}\omega_{4} r^{5}+ \frac{1}{3} C_{5}r^{3}+C_{6},\label{eq25}
\end{equation}
with
\[
-\frac{2}{5} \left(2+\frac{3}{2}\omega+\frac{1}{2\omega}\right)=\omega_{4}~~
 \]
and
 \[
\left(\frac{27}{20}+\frac{15}{28\omega}+\frac{261\omega}{140}+\frac{3}{35\omega^2}+\frac{27\omega^2}{28}\right) = \omega_{5},
 \]
where $C_i$ for $i$ = 1 to 6 are constants of integrations. Now, at centre we will have $m(0)=0$ for which $m_{i}(0)$ should be zero for $i$= 1, 2, 3. 

Using the above conditions we get $C_{1}=C_{4}=C_{6}=0$. For $\epsilon=1$ we shall get the desired approximate analytic mass solution, so that
\begin{equation}
m \approx m_{0}+ m_{1}+m_{2},\label{eq26}
\end{equation}
and finally
\begin{equation}
 m(r) = a_{1}r^{3}+ a_{2}r^{5}+a_{3}r^{7}, \label{eq27}
\end{equation}
with
 \[
 a_{1}= \left(C_{2}+ \frac{C_{3}}{3} + \frac{C_{5}}{3}\right),
 \]
\[
 a_{2}= C_2^{2}\omega_{3} + C_{2} C_{3}\omega_{4},
\]
\[
a_{3}=C_{2}^{3} \omega_{5}.
\]

To measure the accuracy of the approximate analytic mass solution we now define residual of mass as 
\begin{equation}
Res[m]= m^{\prime} - \frac{1}{2}m^{\prime\prime}r  + m^{\prime\prime}m -\omega_{1} \frac{mm^{\prime}}{r} -
\omega_{2}(m^{\prime})^{2}, \label{eq27a}
\end{equation} 
which can be written as the function of radial distance 
\begin{equation}
Res[m]= \sum_{i=2}^{6} \zeta_{i}(n,\omega) r^{2i}, \label{eq27b}
\end{equation}
where $\zeta_{i}(n,\omega)$ are the coefficients. For an exact solution of mass $Res[m]$ will be zero. It is possible to get the residual of mass close to zero with proper choice of the parameters $(n,\omega)$. The accuracy of the approximate solution of mass 
will be high if the residual of mass be small as compared to the mass function value. 

Though our sole motivation is to study NS properties using approximate solution of mass, we note that $\omega=-1$  is interesting for two reasons. Firstly, this is related to the existence as well as definite role of the dark energy. Secondly, for $\omega=-1$  the mass function becomes $m=a_{1}r^{3}$ which is the exact solution of the TOV equation (Eq. (14)) where $\omega_{3}$, $\omega_{4}$, $\omega_{5}$ all are zero. Hence the density becomes constant ($\rho=\frac{3a_{1}}{4\pi}$). Again, the pressure becomes constant as follows from the EOS (Eq. (3)). Therefore we get the solution for constant density sphere which is however not physically admissible. So our model valid for $\omega\neq -1$ which does not give any dark energy solution. Therefore, in the next section we find allowed range for $\omega$ to study properties of  neutron star. In addition to this we also note that the exact solution $m=a_{1}r^{3}$ is comprehensively unable to study feasible features of stellar model for $\omega=-1$. This indicates that approximate solution could be interesting to adopt for exploring interesting properties of NS system.

\section{Star modelling}

\subsection{Causality and stability condition}
The speed of sound is defined in the perfect fluid system as
\begin{equation}
v= \sqrt{\frac{dp}{d\rho}}= \sqrt{\omega}.\label{eq28}
\end{equation}
Since the speed of sound must be less than light speed we always have $0 < v^{2} < 1$ for a solution to be causal~\cite{Herrera1992}.
The adiabatic index for isotropic stellar structure is defined as
\begin{equation}
\Gamma= \left(\frac{\rho+p}{p}\right)\frac{dp}{d\rho}= (\omega +1). \label{eq29}
\end{equation}

The stellar structure will be stable if $\Gamma> \frac{4}{3}$~\cite{Chandrashekhar1964,Bondi1964,Wald1984}. Therefore, we can choose a range of $\omega$ to be $\frac{1}{3} <\omega < 1$ on physical consideration that the range confirms that the stellar model satisfies both the causal and stable conditions as stated above. For this chosen range of $\omega$, we always have $\omega_1$, $\omega_2$, $\omega_5$ to be positive and $\omega_3$, $\omega_4$ to be negative.
 
\subsection{Density function and constants of integrations}
Density function of the star can be written using Eqs.(\ref{eq4}) and (\ref{eq27})
\begin{equation}
\rho(r)= \frac{1}{4\pi} ( 3a_{1} + 5 a_{2} r^{2} + 7 a_{3} r^{4} ). \label{eq32} 
\end{equation}
At the centre of star therefore the density becomes 
\begin{equation}
\rho_c=\frac{3}{4\pi}\left(C_{2}+ \frac{C_{3}}{3} + \frac{C_{5}}{3}\right).\label{eq33}
\end{equation}
The value of the constant of integration is  in general arbitrary. But not all the values of the constants are physically valid. So, some particular values of these constants are interesting as they play a vital role to determine the central density of a stable stellar system. Therefore, without loss of generality to constrain the value of constants for physically interesting star we can introduce a model parameter $n$ such that 
\[
C_{2}=\frac{2C_{3}}{(n-3)} = \frac{2C_{5}}{(n-3)}=\frac{4\pi\rho_{c}}{n},
\]
with $n\geq3$. The above equality holds true for  Eq. (\ref{eq33}). It will help us to find the arbitrary constants of integration in terms of the central density and the model parameters. Therefore, we get the following relationships:

\[
 a_{1}= \frac{C}{3},~a_{2}=\frac{C^{2}}{3n}\omega_3,~a_{3}=\frac{C^{3}}{n^{3}}\omega_5,
 \]
with 
  \[
 4\pi\rho_{c}=C.
 \]

We have now $a_{1}, a_{3}>0$ and $a_{2}<0$ for the features of the causal and stable stellar structure. Eventually we see that there is only one  unknown constant, namely, the central density ($\rho_{c}$) and two parameters, namely, the EOS parameter ($\omega$) and the model parameter ($n$) which can determine different features of  a stellar model. From this model, if the value of the central density is given, by tuning the parameters we can easily determine the total mass and radius for a particular star. Again for star of known mass and radius, tuning the parameters we can find its central density.

\subsection{Pressure, radius and total mass of the star} 
From the TOV equation it is seen that there is a significant contribution of the pressure term in the gravitational force. 
With the help of  Eqs. (\ref{eq3}) and (\ref{eq32}), we get the pressure of star as
\begin{equation}
p(r)= \frac{\omega}{4\pi} ( 3a_{1} + 5 a_{2} r^{2} + 7 a_{3} r^{4} ). \label{eq35}
\end{equation}
The boundary of a star is defined by the relation $p(R)=0$, which gives
\begin{equation}
7\omega_{5}\left(\frac{C}{n}R^{2}\right)^{2}+ \frac{5}{3} n \omega_{3} \left(\frac{C}{n}R^{2} \right)  + n = 0.\label{eq36}
\end{equation}
Solution of  Eq. (\ref{eq36}) is the radius of the star for a given set of values $(n,\omega)$. Hence the radius of the star is given by
\begin{equation}
R^{2}=\frac{n^{2}}{C} \Omega, \label{eq37}
\end{equation}
with $\Omega=\Omega_{0} \pm \Omega_{1}$, where $\Omega_{0} =-\frac{5\omega_{3}}{42\omega_{5}}$ and
$\Omega_{1} =\left(\sqrt{\left(\frac{5\omega_{3}}{42\omega_{5}}\right)^{2}-\frac{1}{7n\omega_{5}}}\right)$.
  
From Eq. (\ref{eq37}), we see that the radius of a star depends on the model parameter, equation of state parameter and central density. It is inversely proportional to square root of central density $(R \propto \rho_c^{-1/2})$. For real values of $R$ we should have                       $\left[\left(\frac{5\omega_{3}}{42\omega_{5}}\right)^{2}-\frac{1}{7n\omega_{5}}\right]\geq 0$ or  $n \geq \left(\frac{252\omega_{5} }{25\omega_{3}^2}\right)$. This is the lower limit for $n$ and it depends on $\omega$. In the range $\frac{1}{3}< \omega < 1$, we have $7.8< \left(\frac{252\omega_{5} }{25\omega_{3}^2}\right)<8.4$. So, for $n<7.8$ there will be no real values for the radius of star for the given range of $\omega$.
 Therefore in our model the minimum value of $n$ is  7.8 which depends on some particular value of $\omega$. However if $n>8.4$ we can ensure the real value of radius of star for given range of $\omega$.
The total mass of the star can be determined evaluating the integral
\begin{equation}
M = 4\pi\int_{0}^{R}\rho(r)r^{2}dr, \label{eq38}
\end{equation}
which reduces to
\begin{equation}
M= \frac{1}{3}CR^{3}+\frac{\omega_{3}}{3n}C^{2}R^{5}+\frac{\omega_{5}}{n^{3}}C^{3}R^{7}.\label{eq39}
\end{equation} 
Rest mass energy of the stellar structure is given by 
\begin{equation}
M_{r} = 4\pi\int_{0}^{R}\rho(r)\left[1-\frac{2m(r)}{r}\right]^{-1/2}r^{2}dr. \label{eq40}
\end{equation}
We can calculate the rest mass energy and total mass to find binding coefficient ($\sigma$) \cite{Fuloria1989} which is defined as
\begin{equation}
\sigma=\left( \frac{M_{r}-M}{M_{r}}\right).\label{eq41}
\end{equation}
It is noted that $\sigma$ should be positive for bound stellar structures.

\subsection{Compactification factor and redshift}
The compactness of a star is measured by the ratio of the Schwarzschild radius $(R_{S}= 2GM/c^{2})$ and the radius of star $(R)$. We define compactification factor of the star as $u=m(r)/r$, which can be expressed as
\begin{equation}
u(r)=a_{1}r^{2}+ a_{2}r^{4}+a_{3}r^{6}. \label{eq42}
\end{equation}
Also the total mass to radius ratio is given by
\begin{equation}
\frac{M}{R}=\frac{n^3\Omega}{3}\left[\frac{1}{n} + \omega_3 \Omega+ 3 \omega_{5} \Omega^{2}\right]. \label{eq43}
\end{equation}
The ratio is independent of the central density of the star but it depends on $n$ in the range $\frac{1}{3}< \omega < 1$. So, $n$ can be think of compactness tuning parameter for given $\omega$. According to the Buchdahl condition~\cite{Buchdahl1959} for a physical stellar model $M/R$ is always less than 4/9.  Later on, it is shown that considering the first order phase transition for NS the radius can be given as $R\geq 2.94GM/c^{2}$~\cite{Glendenning1992} which means the compactness will be such that the ratio $M/R$ will be less than 0.34. 
 
Now it can be shown from Eqs. (\ref{eq37}) and (\ref{eq43})  that $M \propto \rho_c^{-1/2}$. This type of relation is found for other analytic solutions like Tolman VII~\cite{Tolman1939}, Nariai IV~\cite{Nariai1950}, Buchdahl~\cite{Buchdahl1967} and Tolman IV (generalized)~\cite{Lake2003} using different ansatz for fixed M/R ratio. But in our model the relation is direct consequence of approximate analytic mass function without considering any preassumed metric potentials and different ans{\"a}tze. The only input to this model is the linear EOS. Lattimer et al.~\cite{Lattimer2005} used this law of variation of the maximum mass and central density as input to set upper limit for the central density of a NS. It is clear that for larger value of observed mass the upper limit for the central density will be lower for the fixed EOS as well as model parameter.

The redshift \cite{Ozel2013} in the internal region of a star is defined by 
\begin{equation}
Z(r)+1= \frac{1}{\sqrt{g_{tt}(r)}}.
\end{equation}
So the surface redshift will be in the form of
\begin{equation}
Z_{s}+1= (1-2u(R))^{-\frac{1}{2}}.\label{eq44}
\end{equation}
Now based on the Buchdahl condition~\cite{Buchdahl1959} for a stable stellar configuration the ratio $M/R$ is always less than 4/9 and hence we should always have $Z_{s} \leq 2$. From our model we get 
\begin{equation}
Z_{s}+1= \left[1-\frac{2n^3\Omega}{3}\left(\frac{1}{n} + \omega_3 \Omega+ 3 \omega_{5} \Omega^{2}\right)\right]^{-\frac{1}{2}}.\label{eq45}
\end{equation}
One can note that the surface redshift depends only on the parameters $(n,\omega)$.

\subsection{Time-Time component of the metric}
The time-time component of metric tensor in the line element taken into consideration is unknown. This can be derivable from Eq. (\ref{eq5}) using Eqs. (\ref{eq3}) and (\ref{eq4}). The simple calculation yields  
\begin{equation}
g_{tt}(r) =  g_{tt}(0) \frac{e^{(\omega+1)I(r)}}{\left(1-\frac{2m(r)}{r}\right)^{\omega}}, \label{eq46}
\end{equation}
where 
\[
I(r)=\int \frac{2(a_{1}r+ a_{2}r^{3}+a_{3}r^{5})}{1-2(a_{1}r^{2}+ a_{2}r^{4}+a_{3}r^{6})}dr.
\]

Note that the time-time metric component is non-singular and hence at the centre we must get a $g_{tt}(0)$ which is finite. The value of $g_{tt}(0)$ can be found by equating the line element to the Schwarzschild metric at the boundary. Thus 
\begin{equation}
g_{tt}(R)= \left(1- \frac{2M}{R}\right). \label{eq47}
\end{equation}
Using this matching condition we can find
\begin{equation}
g_{tt}(0)= \left(1- \frac{2M}{R}\right)^{\omega+1}e^{-(\omega+1)I(R)}.\label{eq48}
\end{equation}

\subsection{Energy condition} 
A physical stellar model should follow the energy relations as prescribed below:\\
(i)~NEC: $\rho+p\geq 0$,~or~$(\omega+1)\rho\geq 0$.\\
(ii)~WEC: $\rho+p\geq 0,~\rho\geq 0$,~or~$(\omega+1)\rho\geq 0,~\rho\geq 0$.\\
(iii)~SEC: $\rho+p\geq 0,~\rho+3p\geq 0$,~or~$(\omega+1)\rho\geq 0,~(3\omega+1)\rho\geq 0$.\\
(iv)~DEC: $\rho\geq 0,~\rho> \mid p \mid$.

In our model if the density profile $\rho\geq0$ be positive throughout the whole region of a star then all the energy conditions will be satisfied.

\subsection{Core and crust structure}
The variation of density profile with the radial distance  will not be similar at the core and crust region of a NS. The core in general is in the supranuclear density regime whereas the crust is in the subnuclear density regime. The regime switches at the point which defines core-crust boundary. In our model the core-crust boundary is at a point where slope of density profile switches its nature of variation. So, extrema of the slope of density curve determines the distance of the core-curst boundary from the centre of NS.

The slope of the density is given by 
\begin{equation}
\rho^\prime(r)= \frac{1}{4\pi}(10 a_{2}r+28a_{3}r^{3}).\label{eq49}
\end{equation}
Let the extrema of the slope of density is found to be at $r=r_{cc}$. So, we have  
\begin{equation}
r^{2}_{cc}=-\frac{10a_{2}}{84a_{3}}=-\frac{5n^{2}\omega_{3}}{126C\omega_{5}}.\label{eq50}
\end{equation}
The radius of the star and $r_{cc}$ are related by the following relation
\begin{equation}
R^{2}= 3 r^{2}_{cc} \pm \sqrt{9r^{4}_{cc}-\frac{n^{3}}{7\omega_{5}C^{2}}} .\label{eq51}
\end{equation}
The density at core-crust boundary is related to the central density as 
\begin{equation}
\rho_{cc}=\rho_{c}\left(1-\frac{125n\omega_{3}^{2}}{2268\omega_{5}}\right).\label{eq51a}
\end{equation}

\section{Some special solutions for neutron stars}
If the central density is specified, for a given set of values of ($n$, $\omega$) we shall have the following three solutions set for the model. 

\subsection{Solution I:} 
\begin{figure}[!htp]\centering
\includegraphics[width=6cm]{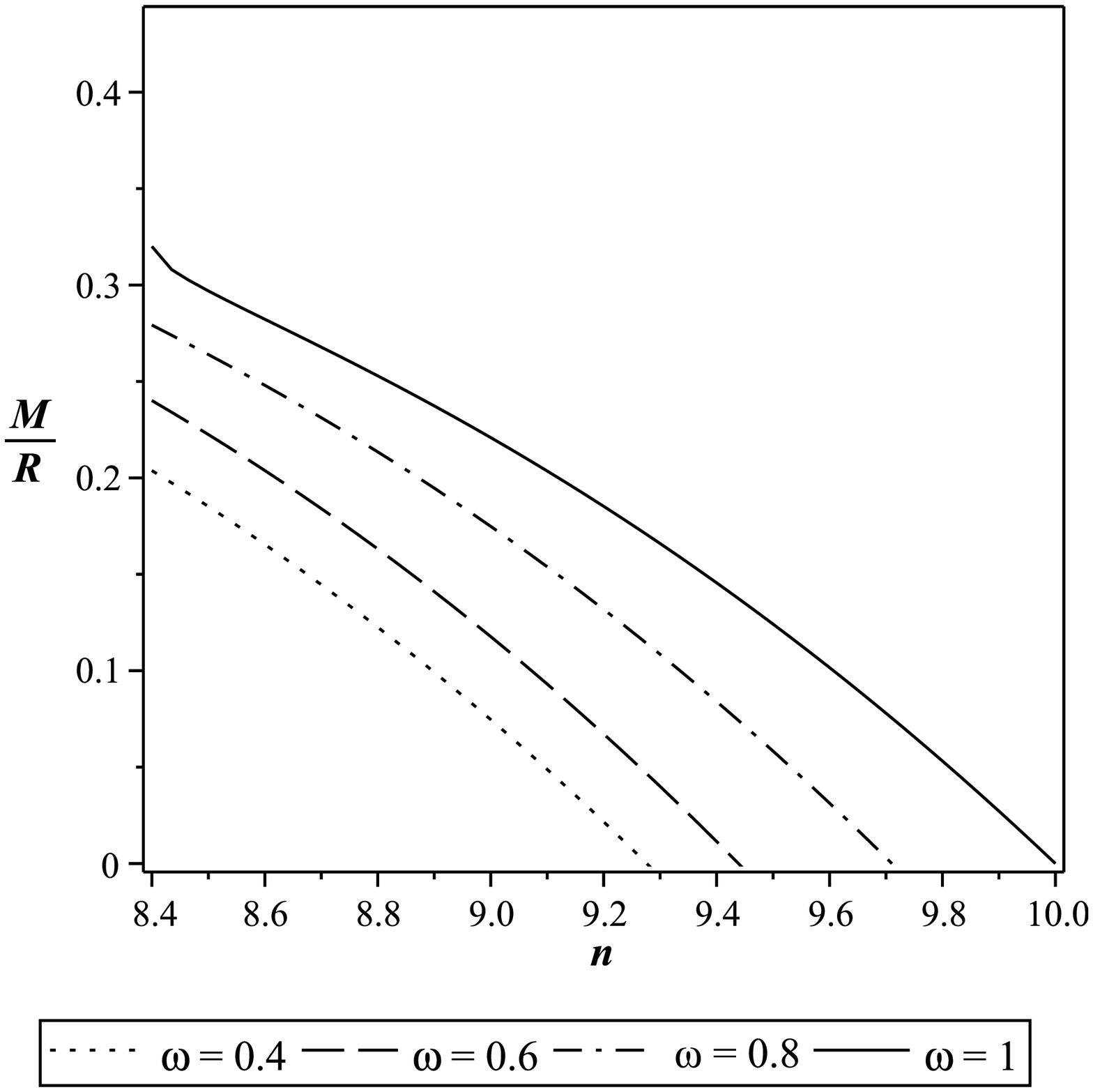} 
\includegraphics[width=6cm]{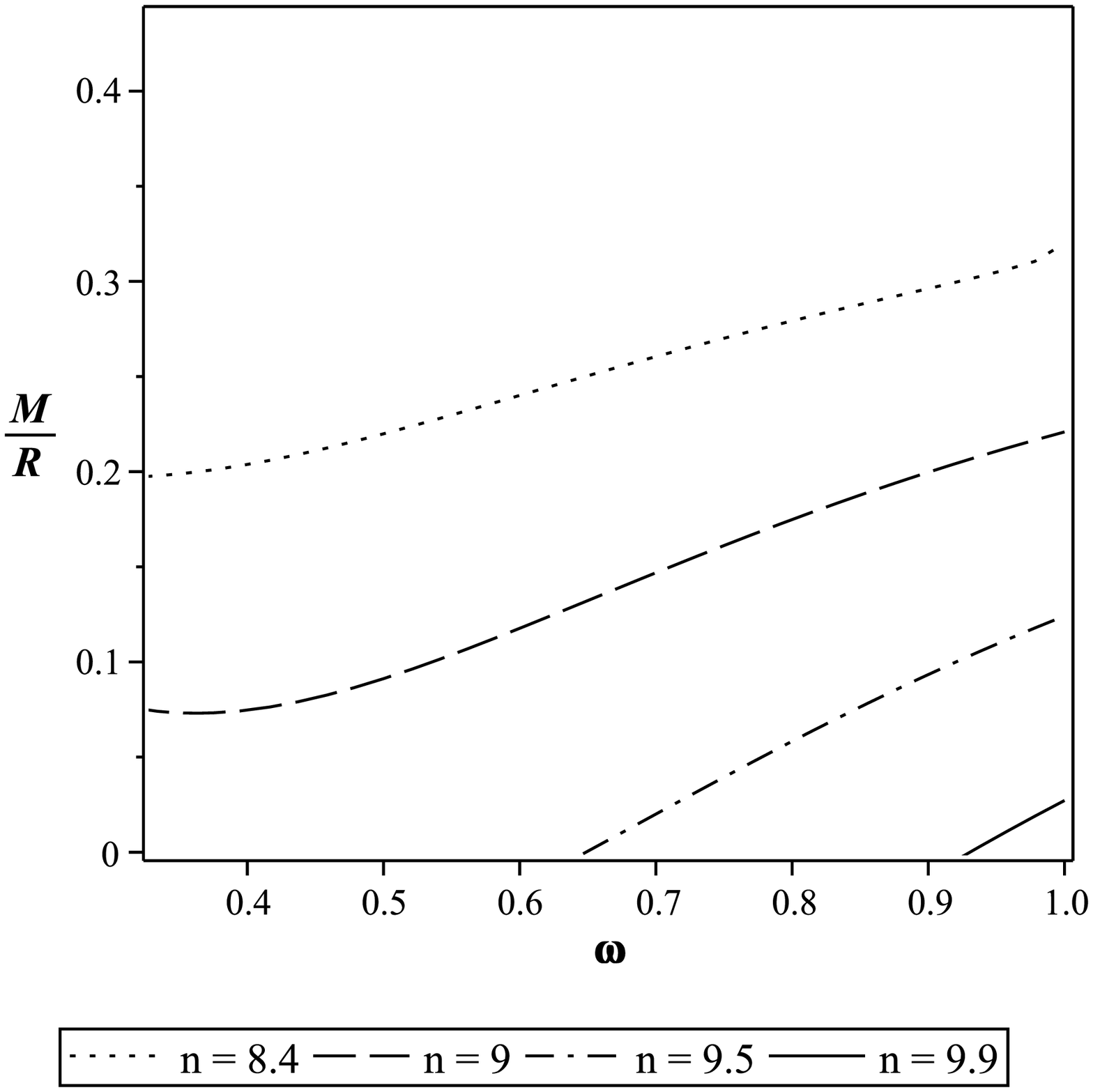}
\includegraphics[width=6cm]{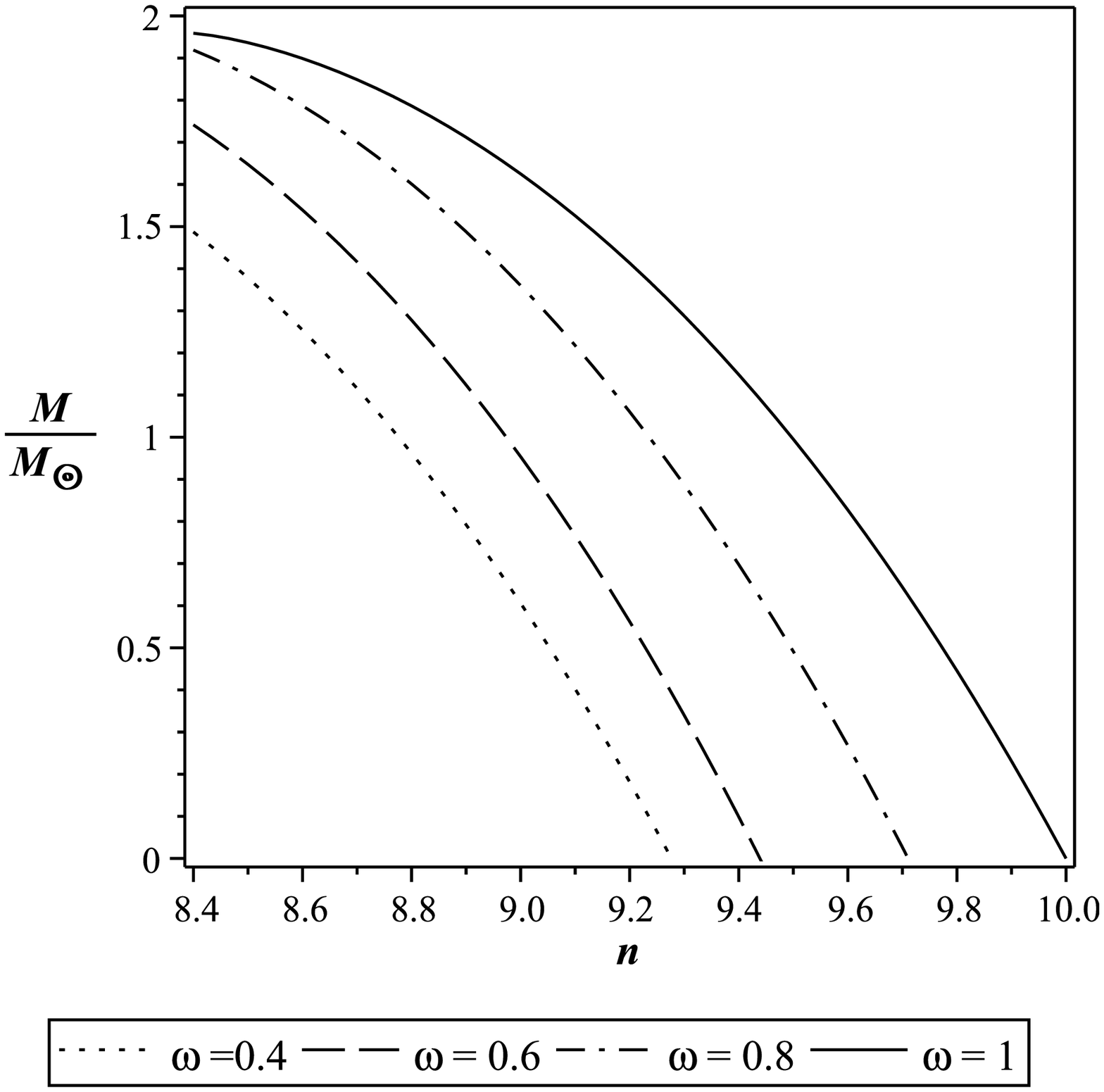} 
\includegraphics[width=6cm]{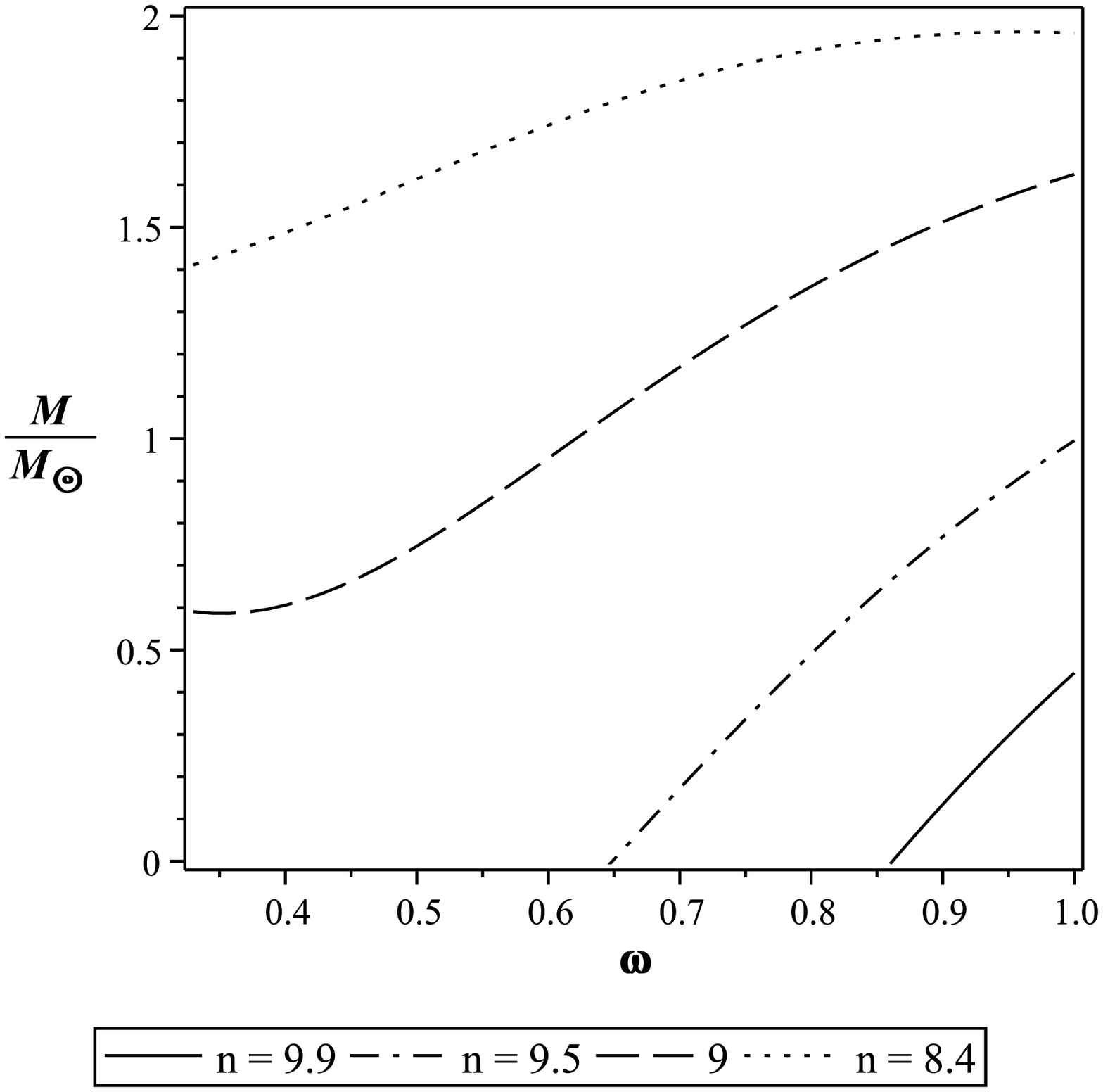} 
\includegraphics[width=6cm]{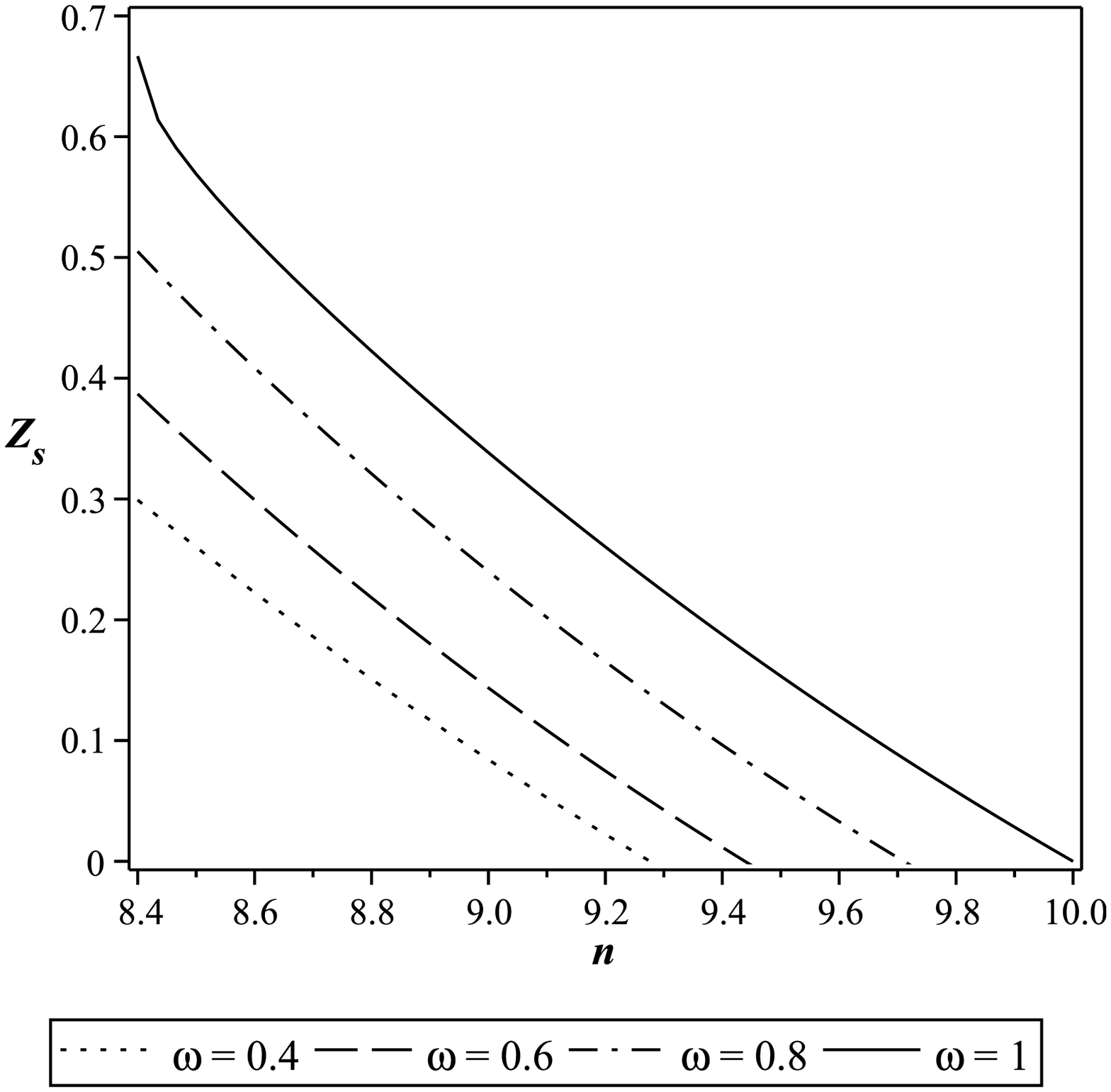}
\includegraphics[width=6cm]{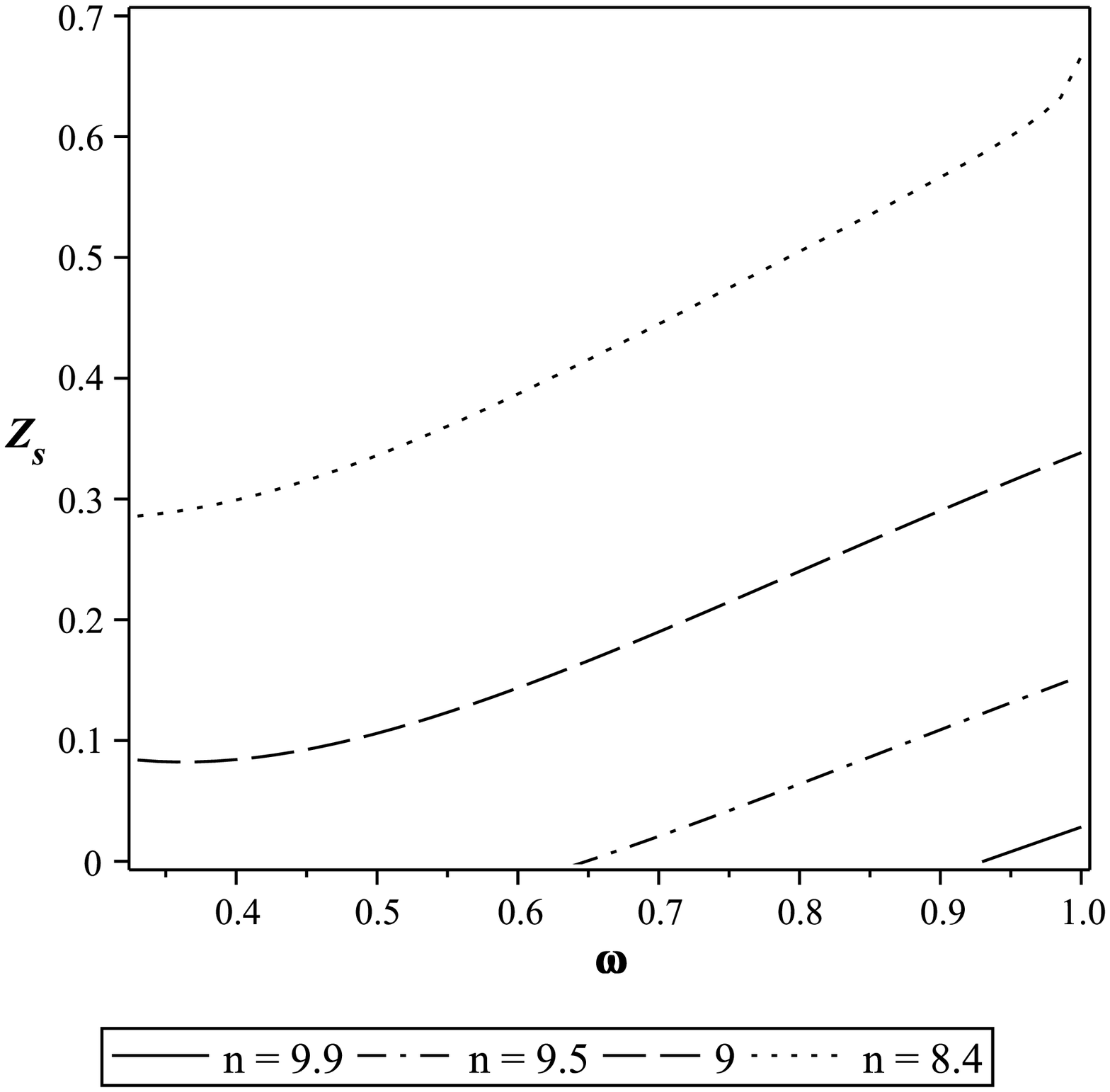}
	\caption{Variation of the total mass to the radius ratio (M/R), the total mass (M) at the central density $5.5\times10^{15}~ gm/cm^{3}$ and the surface redshift ($Z_s$) with respect to the EOS parameter $\omega$ and the model parameter $n$ in the case of Solution I.}
	\label{Fig1}
\end{figure}

If we consider $\Omega=(\Omega_{0} + \Omega_{1} $), we get the following results from Fig. \ref{Fig1}.
\\
(i) The ratio $M/R$ is decreasing with increasing value of $n$. The allowed ranges are $0\leq \frac{M}{R}\leq 0.32$ and  $8.4\leq n \leq10$.
\\
(ii) The maximum surface redshift is 0.67 at $n=8.4$ and $\omega \approx 1$.
\\
(iii) For a given set of (n, $\omega$) we always find pressure vanishes at two points, namely  $R_{o} =n \sqrt{\frac{\Omega_{0}+\Omega_{1}}{C}}$ and $R_{i}=n \sqrt{\frac{\Omega_{0}-\Omega_{1}}{C}}$. Thus the star exhibits outer radius $R_{o}$ and inner radius $R_{i}$.  Since $R_{i} < R_{o}$ we find shell of thickness $(R_o - R_i)$ where the density as well as pressure  become negative and also energy conditions violate.  
 \\                                                                                                                                                                                        (iv) For a star of the central density $5.5\times10^{15}~g/cm^{3}$ the maximum mass is 1.96 $M_\odot$ corresponding to the parameters $n=8.4$ and $\omega \approx 1$. Here the radius of the star is 9.06 km.   

\subsection{Solution II:} 
\begin{figure}[!htp]\centering
\includegraphics[width=6cm]{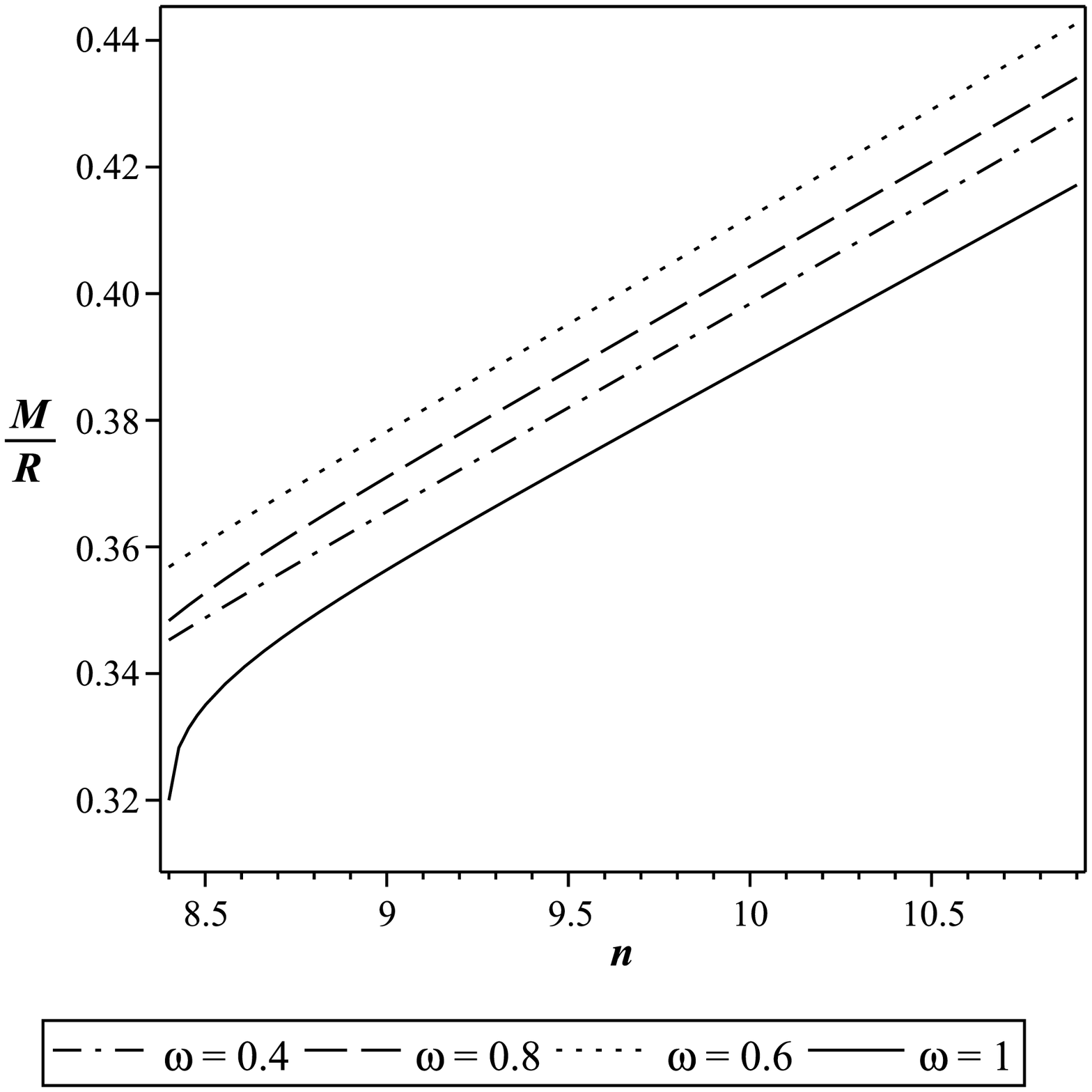} 
\includegraphics[width=6cm]{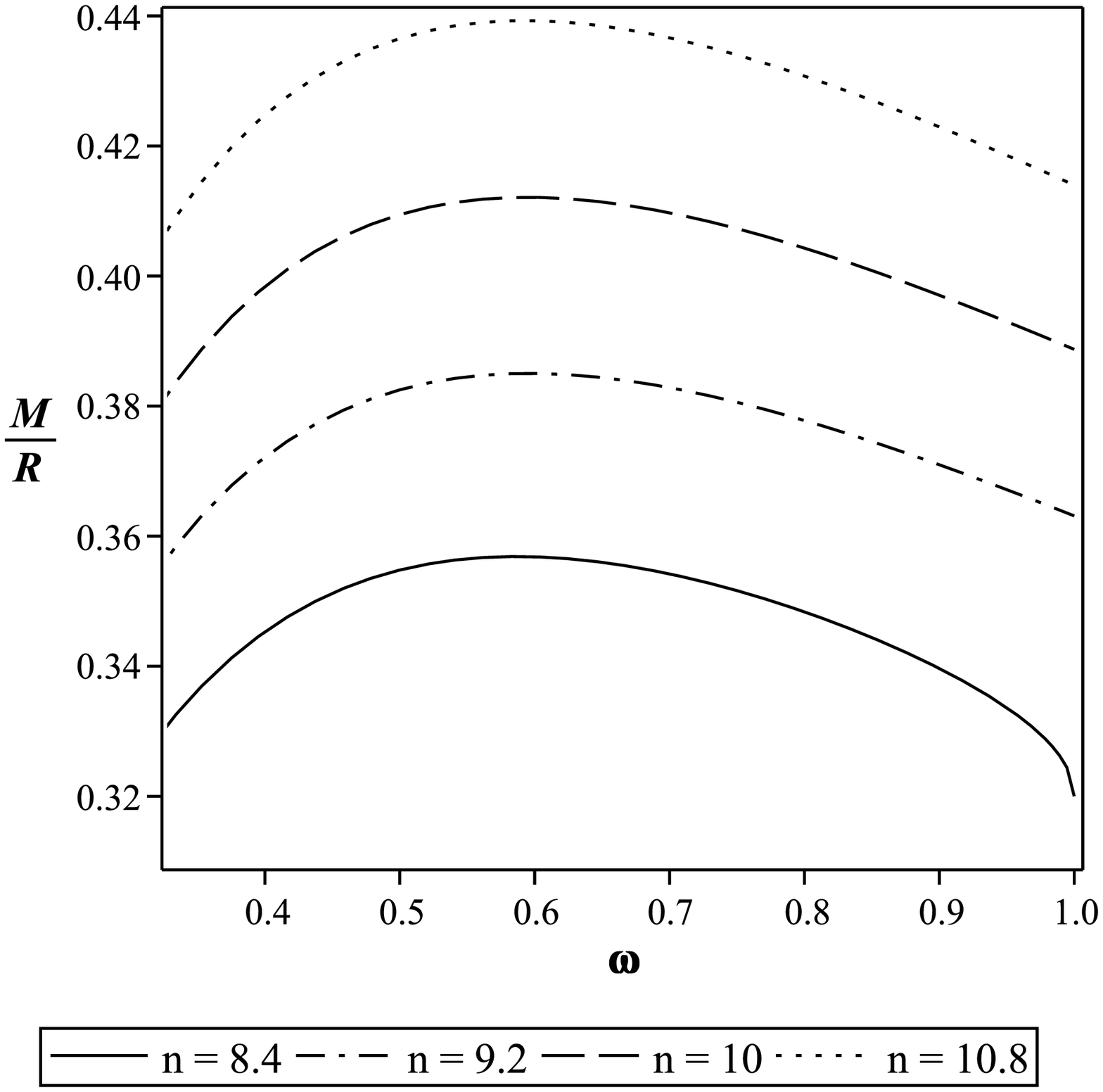}
\includegraphics[width=6cm]{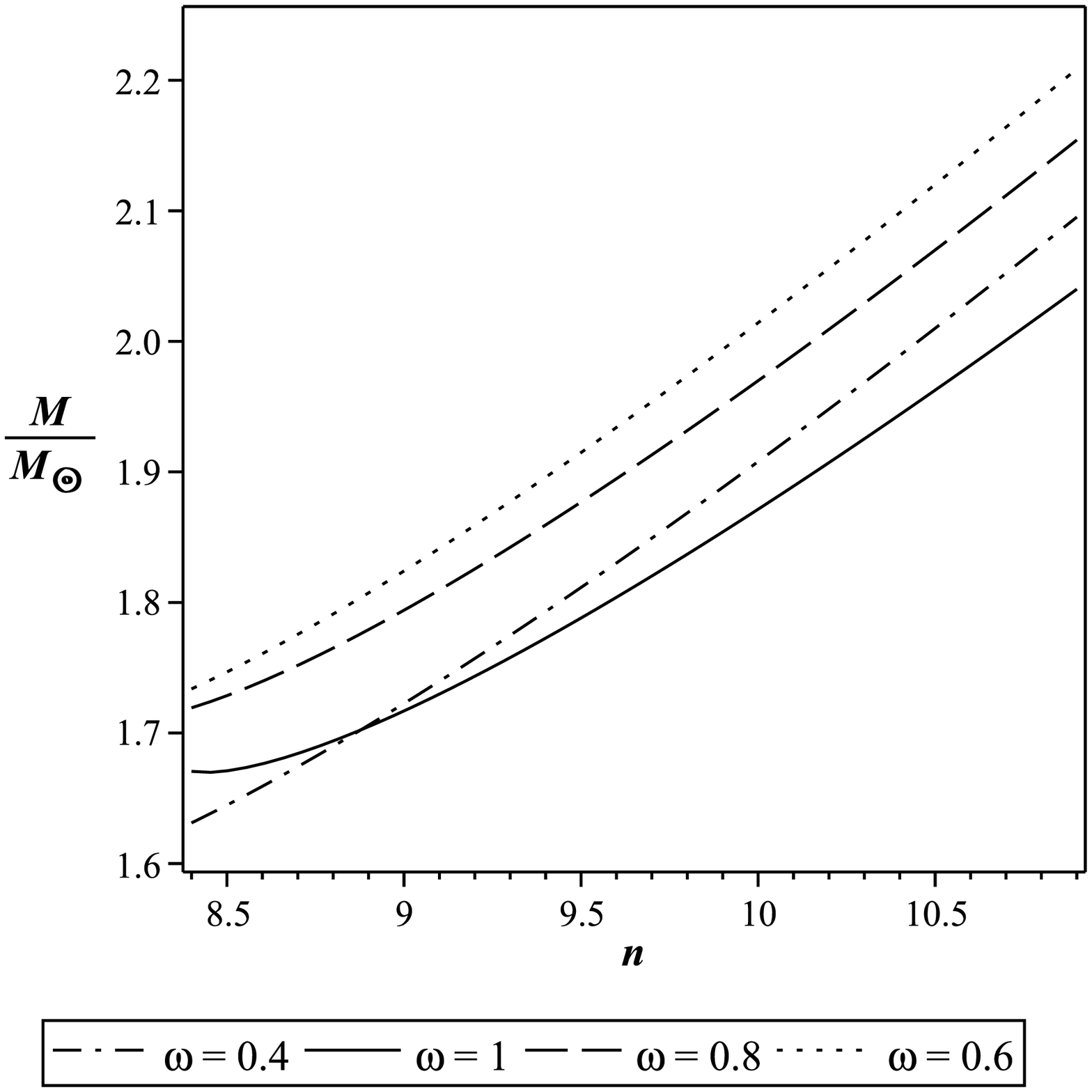} 
\includegraphics[width=6cm]{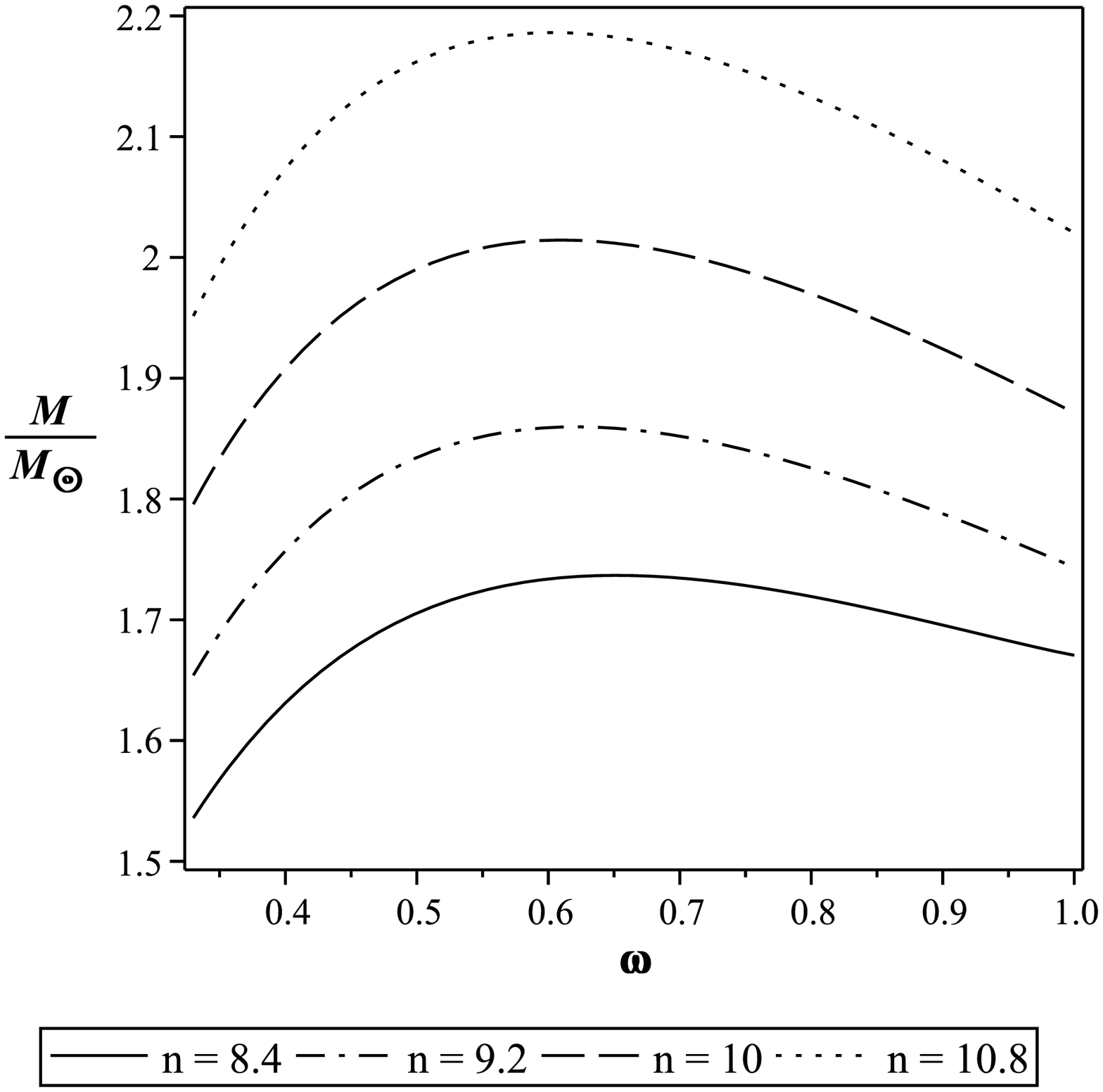} 
\includegraphics[width=6cm]{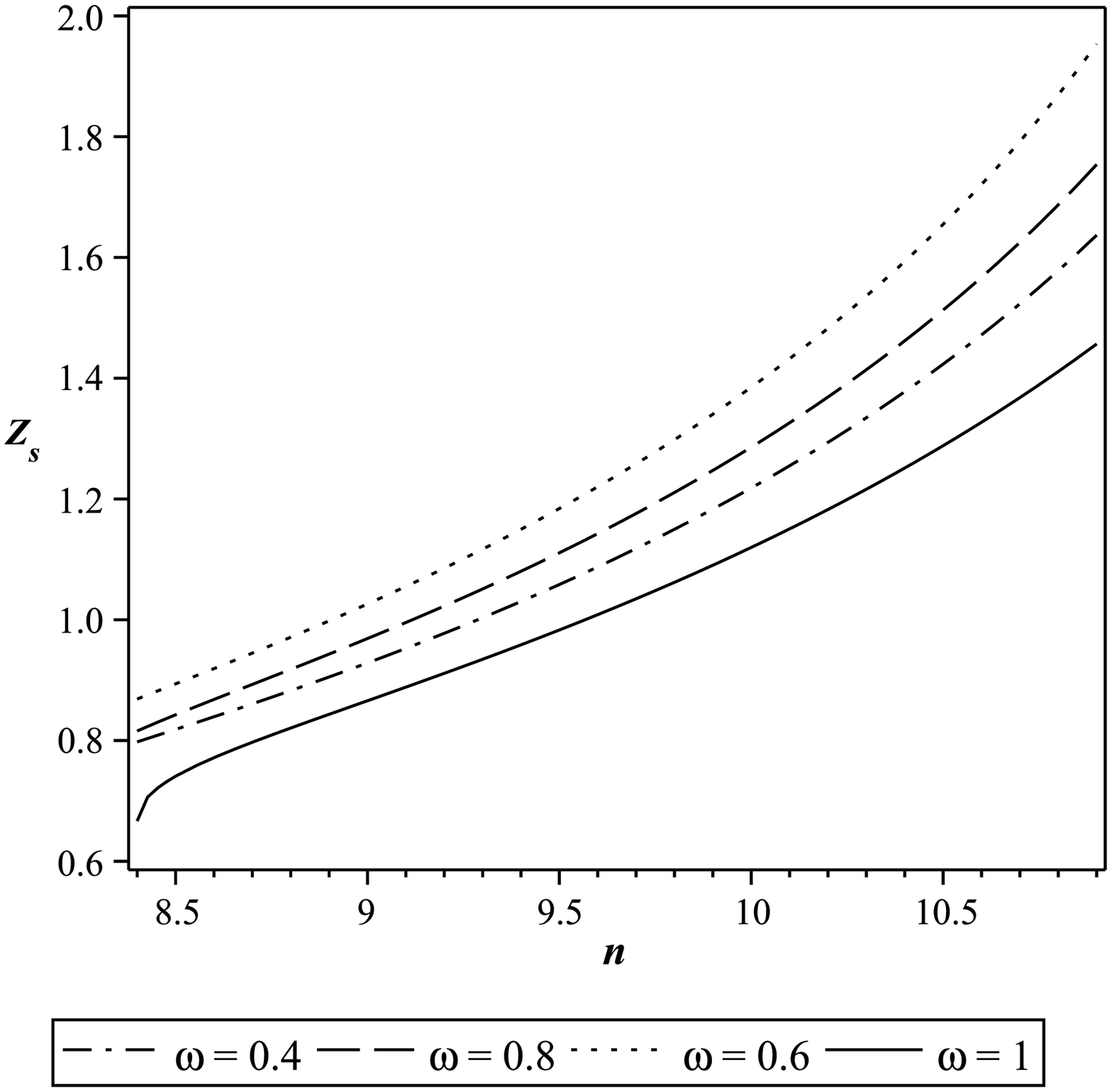}
\includegraphics[width=6cm]{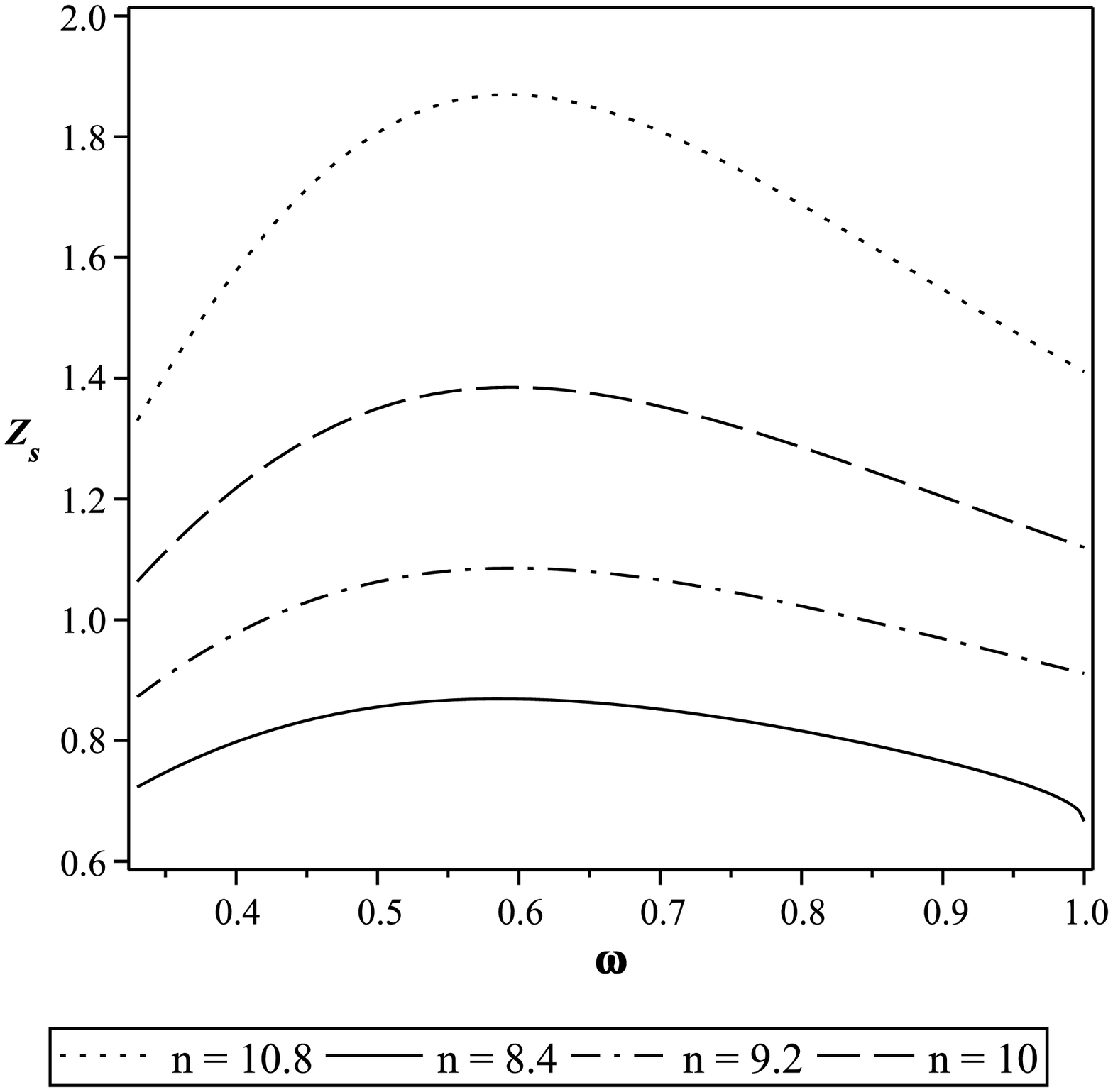} 
	\caption{ Variation of the total mass to the radius ratio (M/R), the total mass (M) at the central density $5.5\times10^{15}~ gm/cm^{3}$ and the surface redshift ($Z_s$) with respect to the EOS parameter $\omega$ and the model parameter $n$ in the case of Solution II.}
	\label{Fig2}
\end{figure}

If we consider $\Omega=(\Omega_{0} -\Omega_{1} $), we get the following results from Fig. \ref{Fig2}.
\\
(i) The ratio $M/R$ is increasing with increasing $n$. The allowed ranges are $0.32\leq \frac{M}{R}\leq 0.44$ and  $8.4\leq n \leq10.9$.
\\
(ii) The maximum surface redshift is 1.95 at $n=10.9$ and $\omega=0.59$. 
\\
(iii) The radius of the star is $R=n \sqrt{\frac{\Omega_{0}-\Omega_{1}}{C}}$. 
\\
(iv) In this range for a star of the central density $5.5\times10^{15}~g/cm^{3}$ the maximum mass is 2.48 $M_\odot$ corresponding to the parameters $n=10.9$ and $\omega=0.59$. The radius of the star is 8.66 km.   

\subsection{Solution III:}  
\begin{figure}[!htp]\centering
\includegraphics[width=6cm]{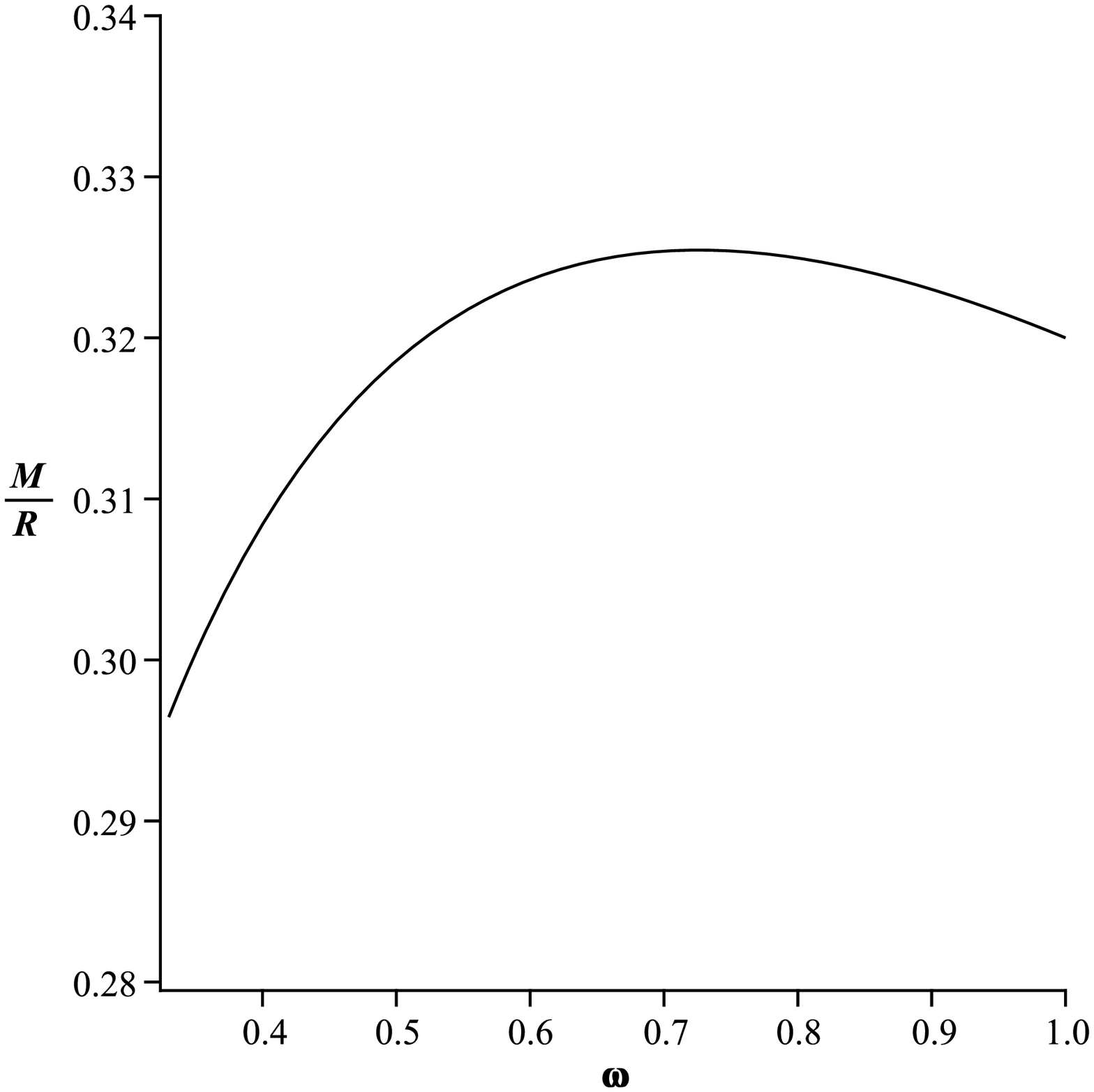} 
\includegraphics[width=6cm]{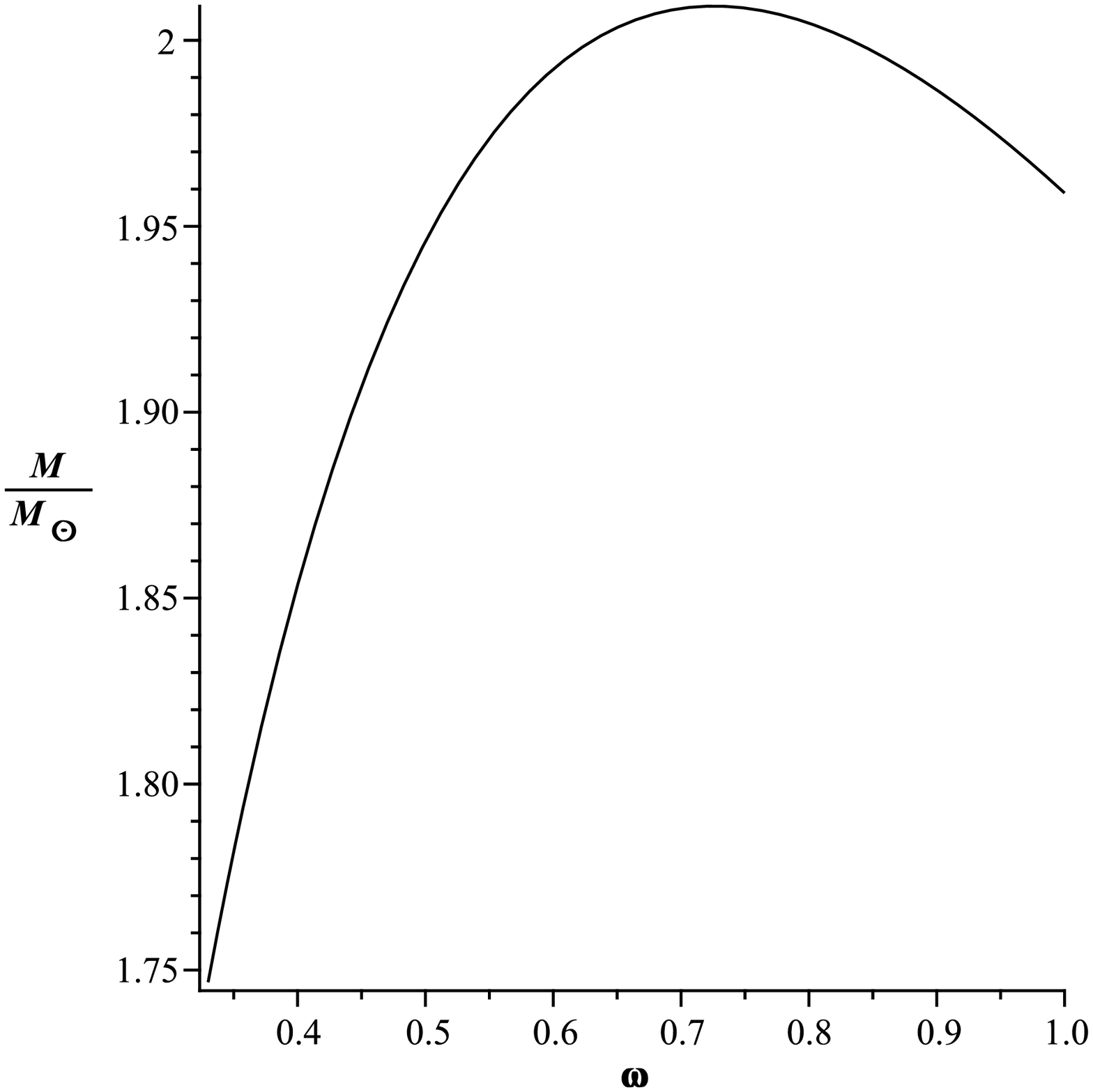}
\includegraphics[width=6cm]{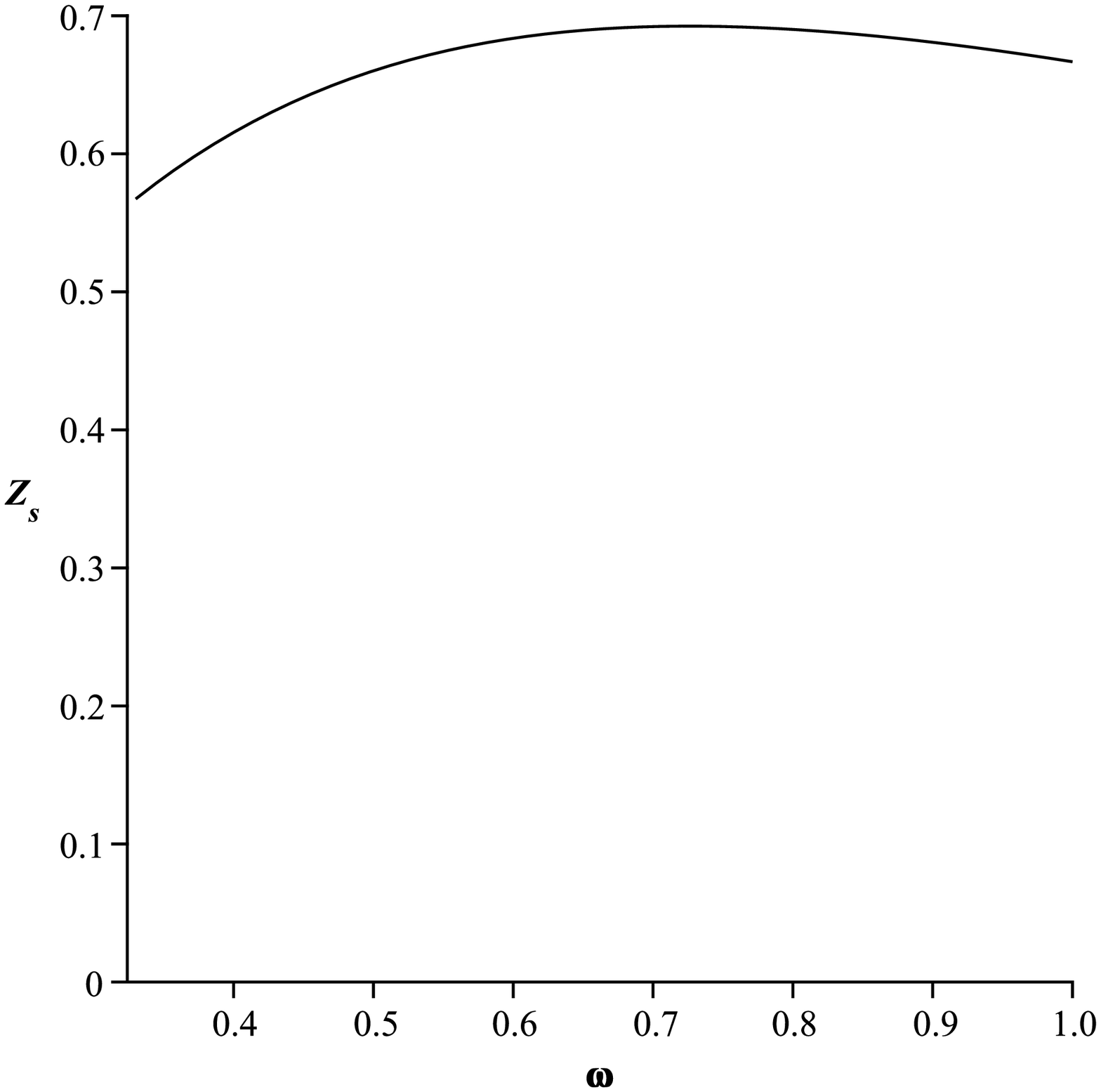} 
	\caption{ Variation of the total mass to the radius ratio (M/R), the total mass (M) at the central density $5.5\times10^{15}~ gm/cm^{3}$ and the surface redshift ($Z_s$) with respect to the EOS parameter $\omega$ in the case of Solution III.}
	\label{Fig3}
\end{figure}

If we consider $n=\left(\frac{252\omega_{5} }{25\omega_{3}^2}\right)$, then we shall have $\Omega_{1}=0$ and $\Omega=\Omega_{0}$. This solution is independent of the parameter $n$ and depends only on $\omega$ of the EOS parameter and central density will give the complete solution. We get the following results from Fig. \ref{Fig3}. 
\\                                                                                                                              
(i) In this case we get $\frac{M}{R}=\left(-\frac{576\omega_{5}}{625\omega_{3}^{3}}\right)$. The allowed range is $0.29714\leq \frac{M}{R} \leq 0.32544$.  
\\
(ii) Surface redshift is maximum 0.69 at  $\omega=0.73$.
\\
(iii) The radius of the star is $R =n\sqrt{\frac{\Omega_{0}}{C}}$ and total mass $M= \frac{6096384}{78125}\left(\frac{\omega_{5} }{\omega_{3}^2}\right)^{3}\frac{\Omega_{0}^{3/2}}{\sqrt{C}}$.
\\
(iv) For a star of the central density $5.5\times10^{15}~g/cm^{3}$ the maximum mass is 2.01 $M_\odot$ with radius 9.13 km at $\omega=0.73$.

The allowed ranges for parameters and the results for all the three solutions are summarized in Tables \ref{tbl-1} and \ref{tbl-2}.

\section{Physical validity of the stellar model}
Though the mass of NS can be measured precisely but there are large uncertainty in the radii measurements. The present stellar model should be tested for its physical validity by studying some observed NSs with known masses. We therefore calculate several physical parameters from the model, such as the radii, central density, central pressure, central redshift and surface redshift of the stars. From Table \ref{tbl-1} we note that the stars described by the three solutions  belong to different ranges of compactness and surface redshift. Keeping this in mind we test the model with observed stars for Solutions I and III. On the other hand, Solution II describes and predicts the existence of  ultra compact neutron star which possess high surface redshift. 

\begin{figure}
\includegraphics[width= 6cm]{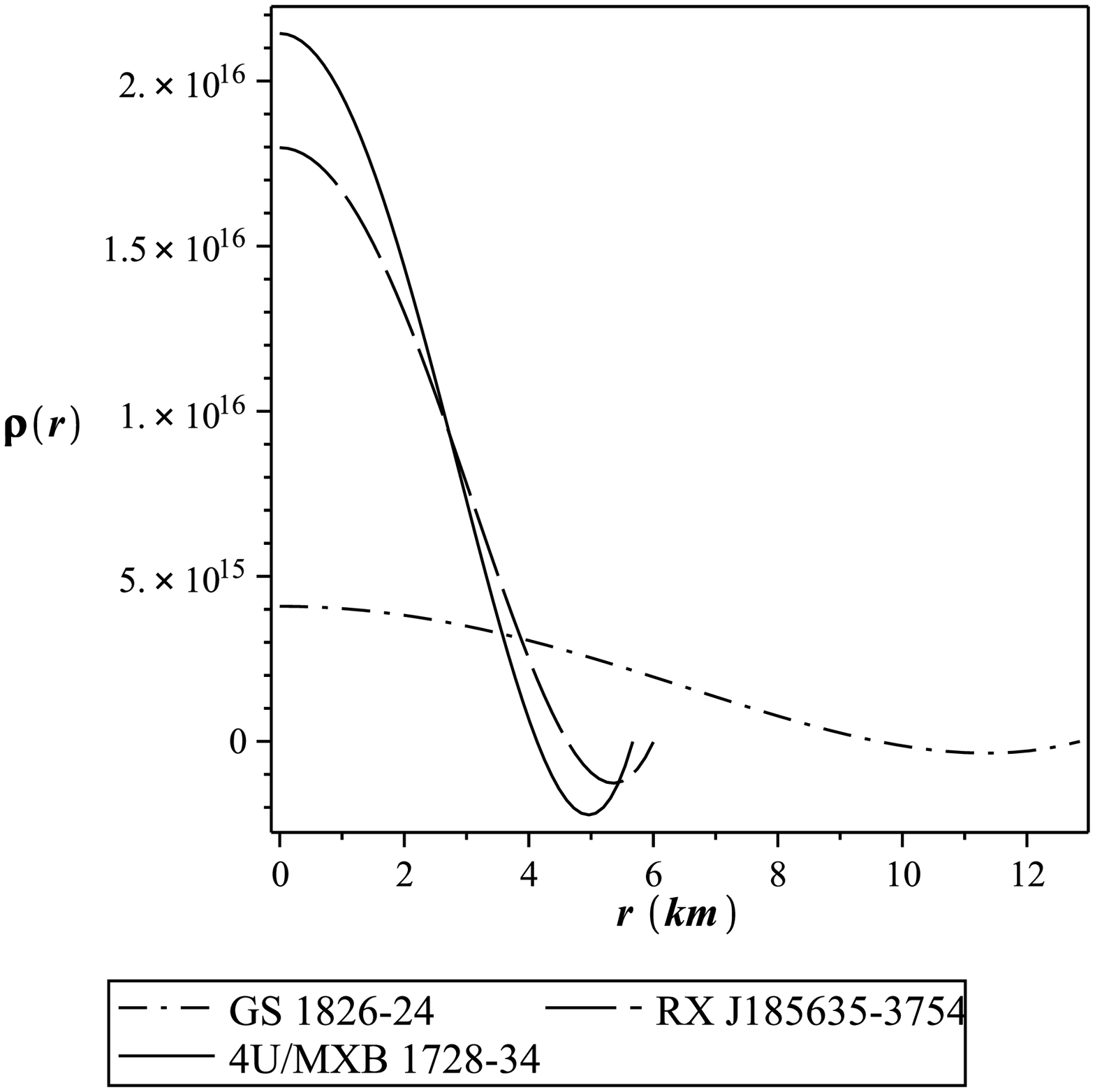}
\includegraphics[width= 6cm]{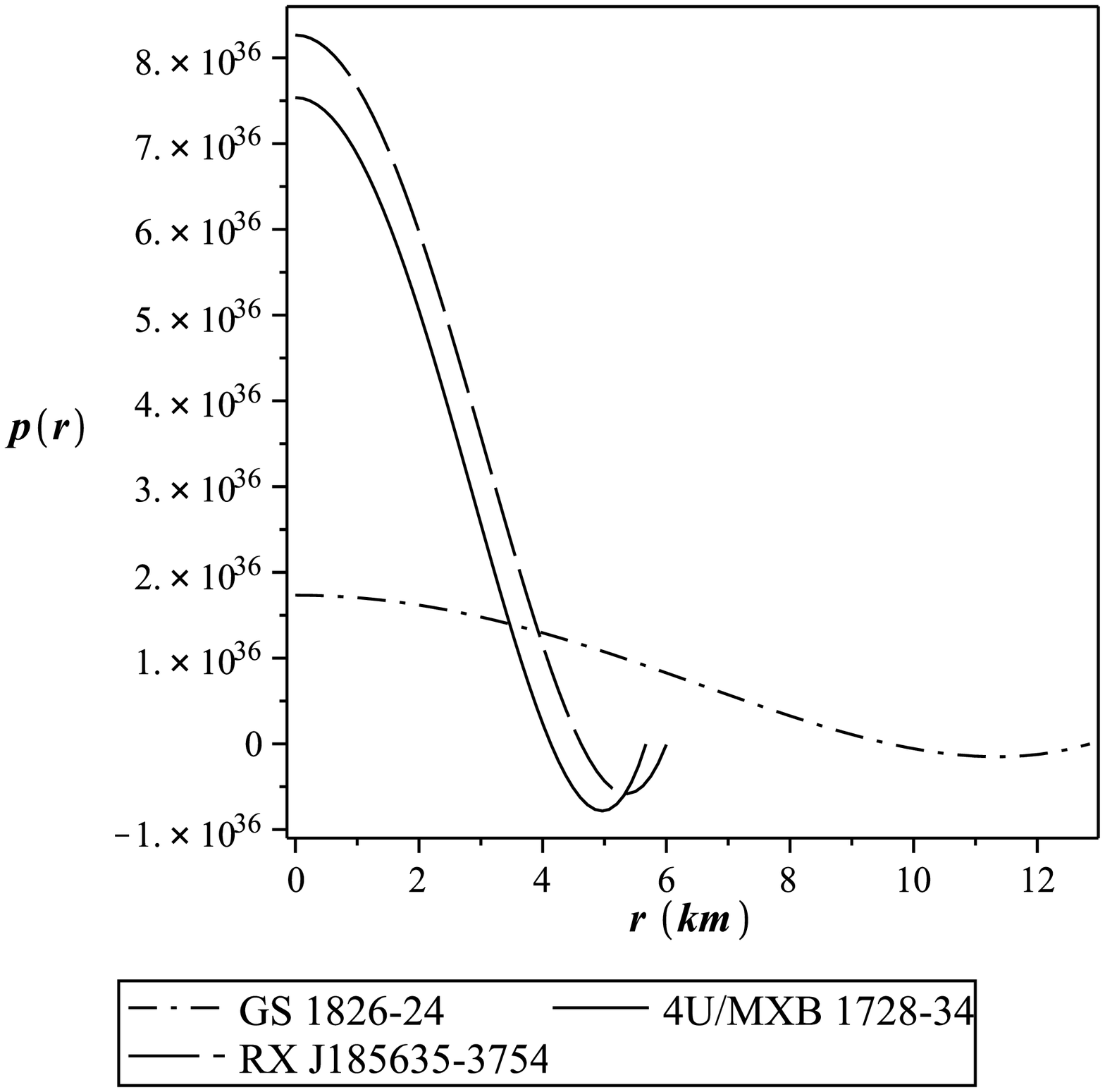}
\centering
\includegraphics[width= 6cm]{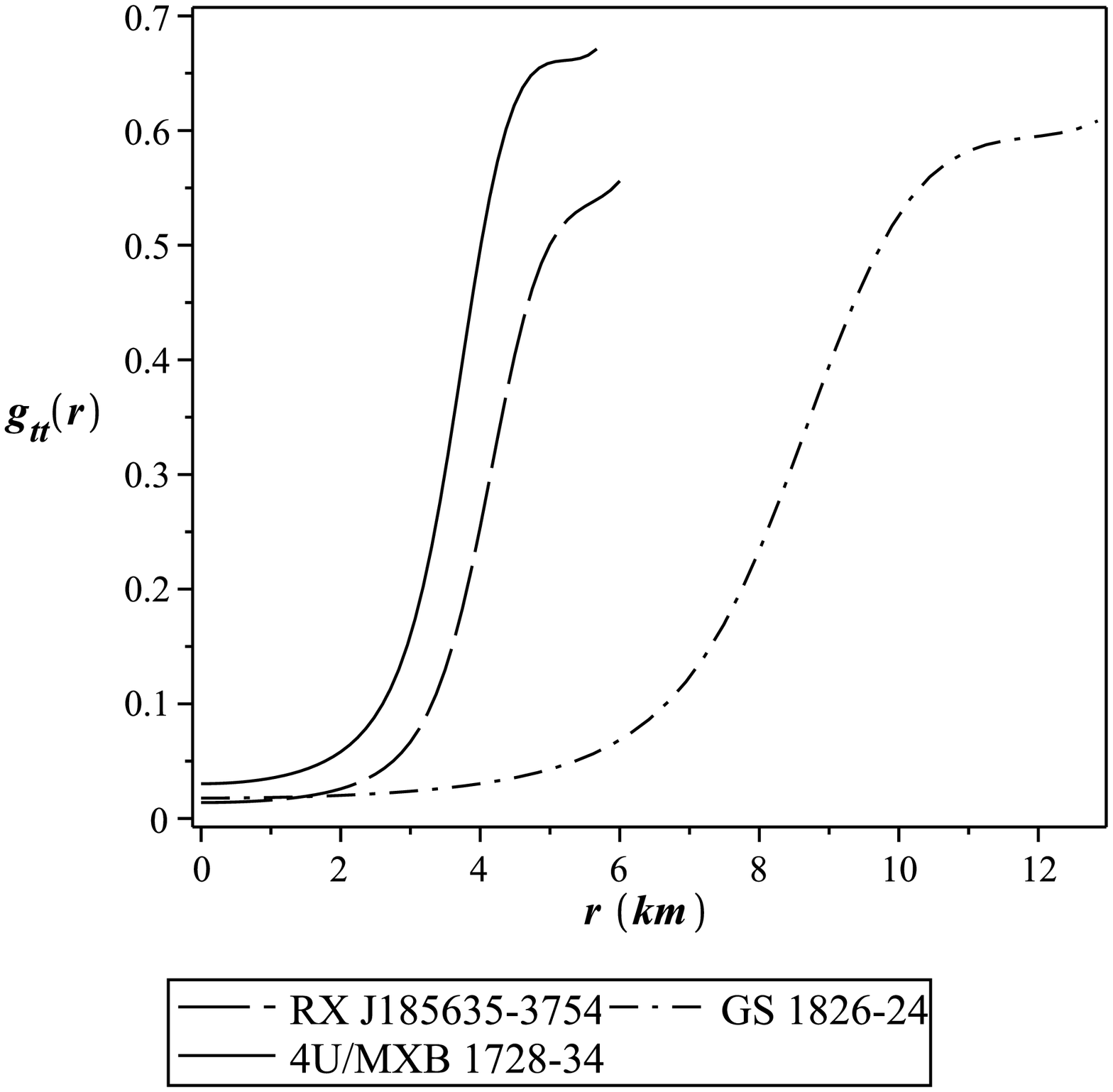}

\includegraphics[width= 6cm]{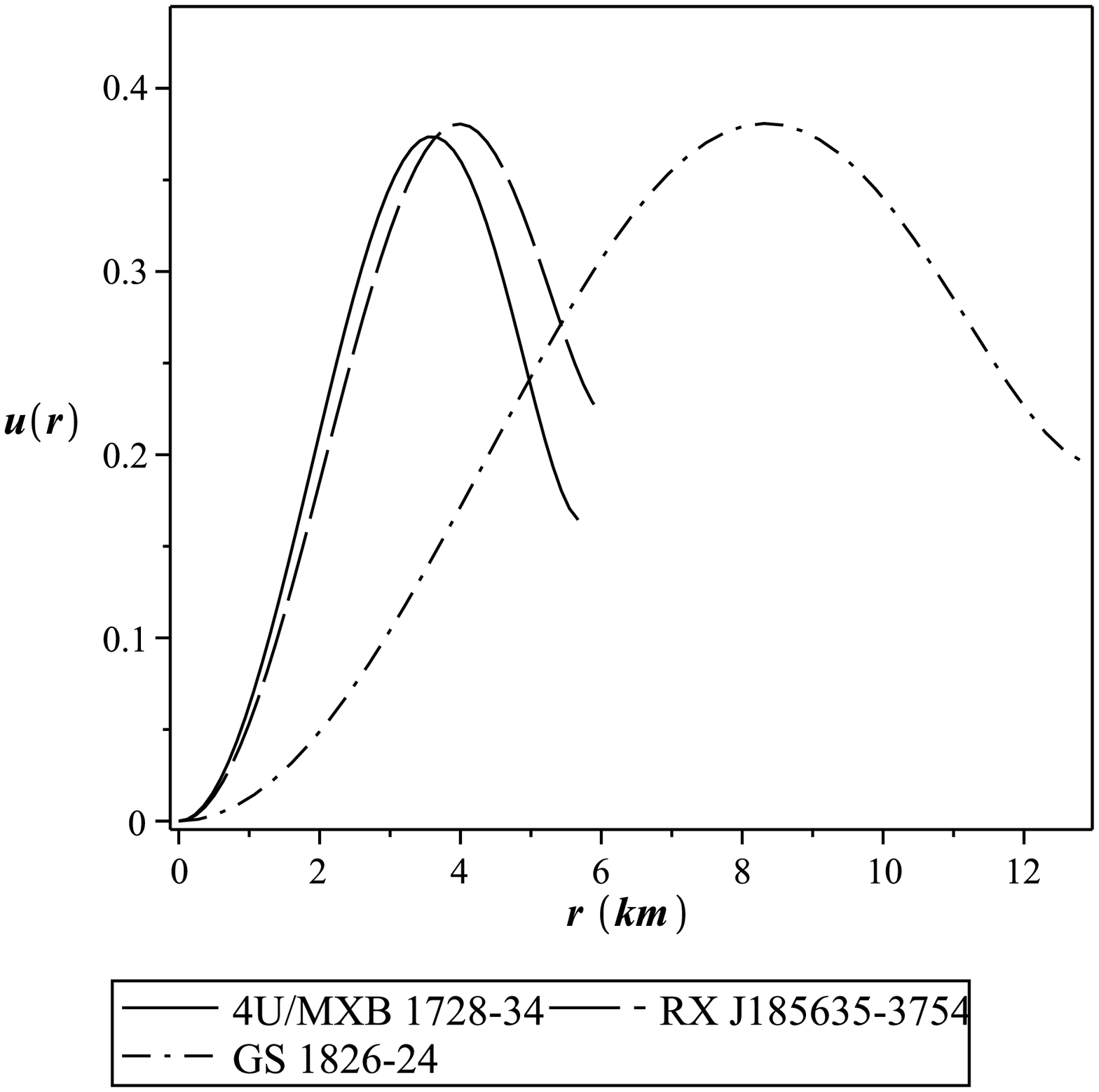}
\includegraphics[width= 6cm]{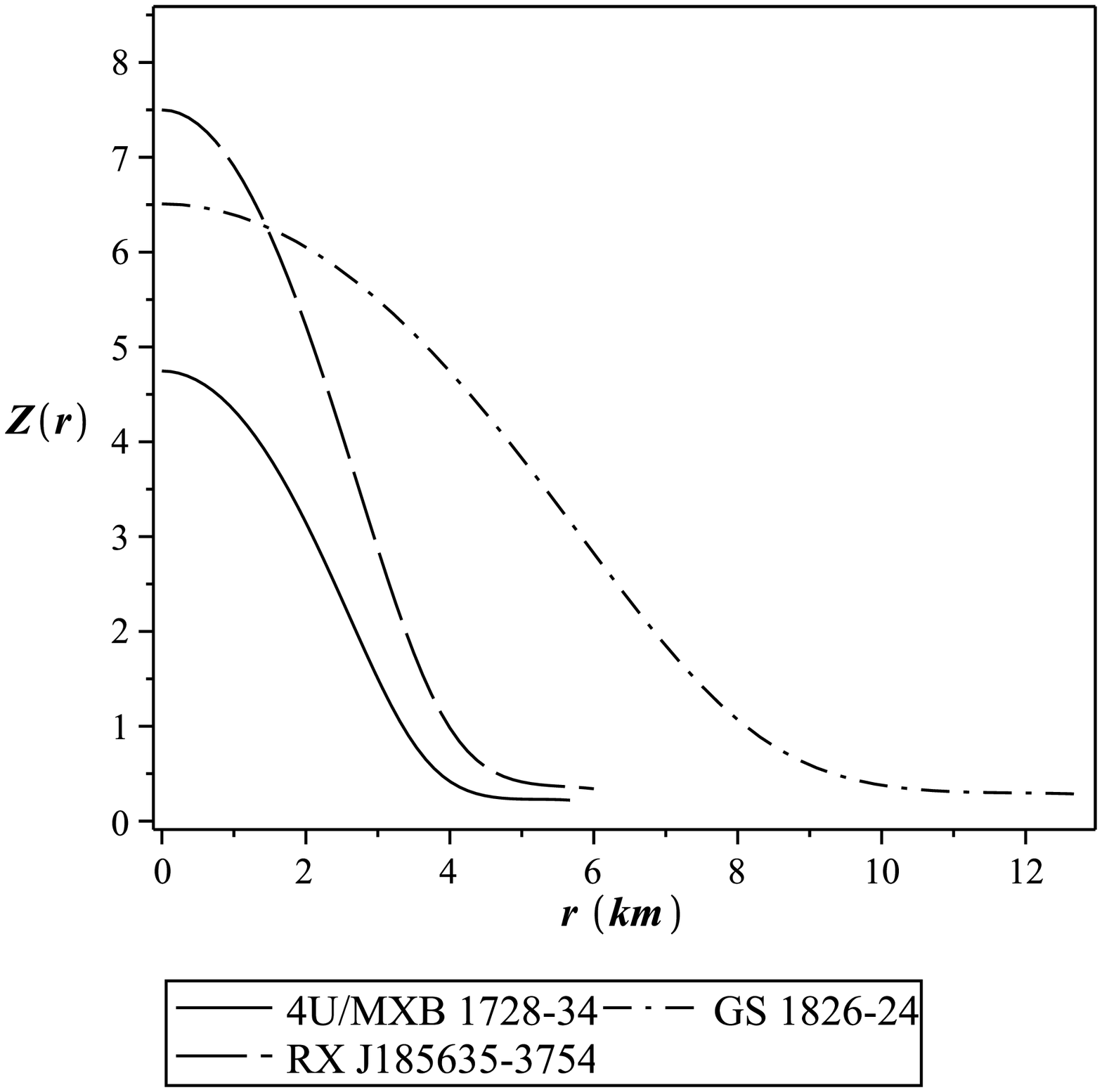}
\caption{Solution I:  Density  in $gm/cm^3$ (upper left), pressure in $dyn/cm^2$ (upper right), the time-time component (middle), compactness factor (lower left) and redshift (lower right) as a function of the radial distance $r$.}\label{Fig4}
\end{figure}

\begin{figure}
\centering
\includegraphics[width= 6cm]{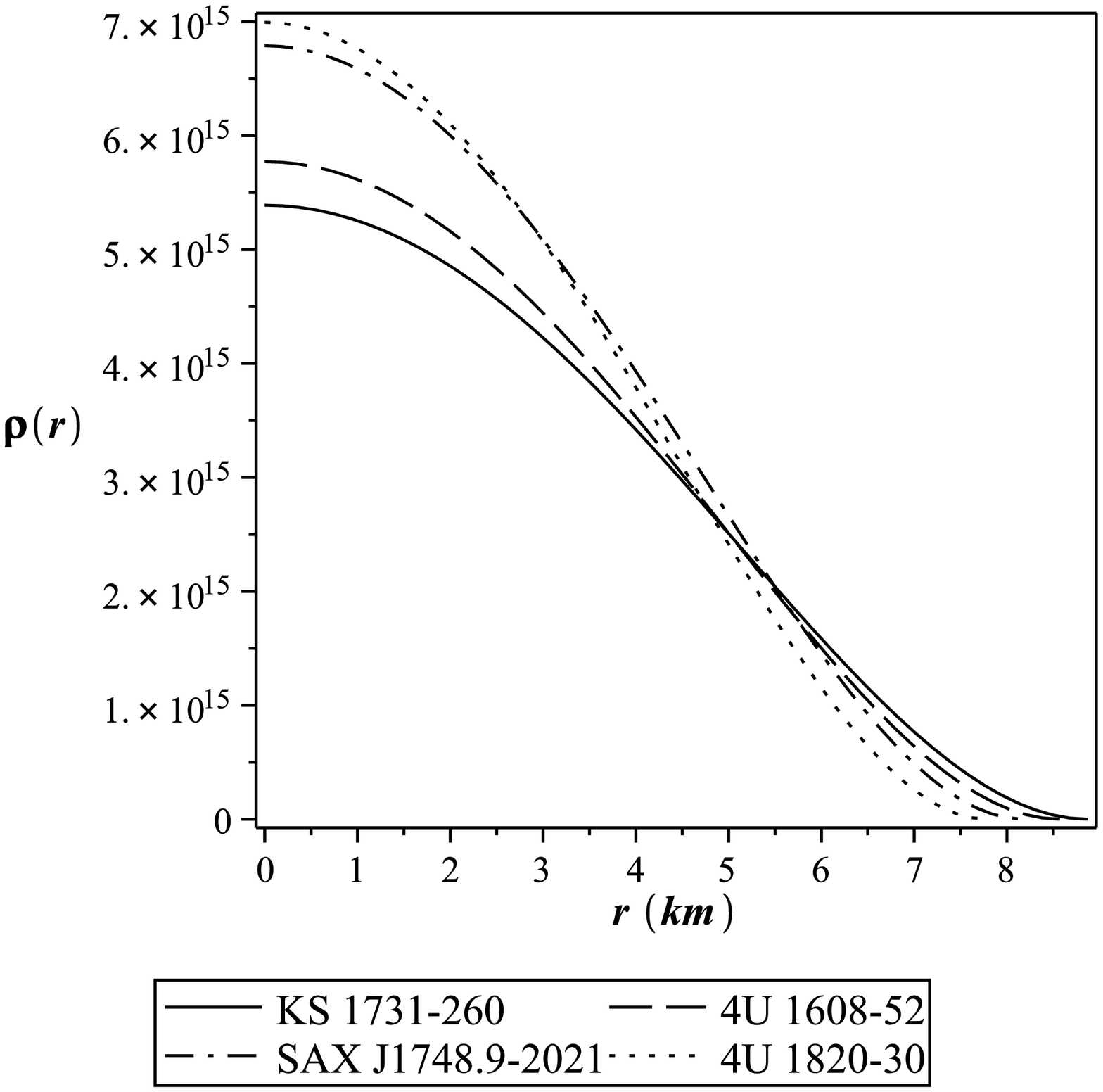}
\includegraphics[width= 6cm]{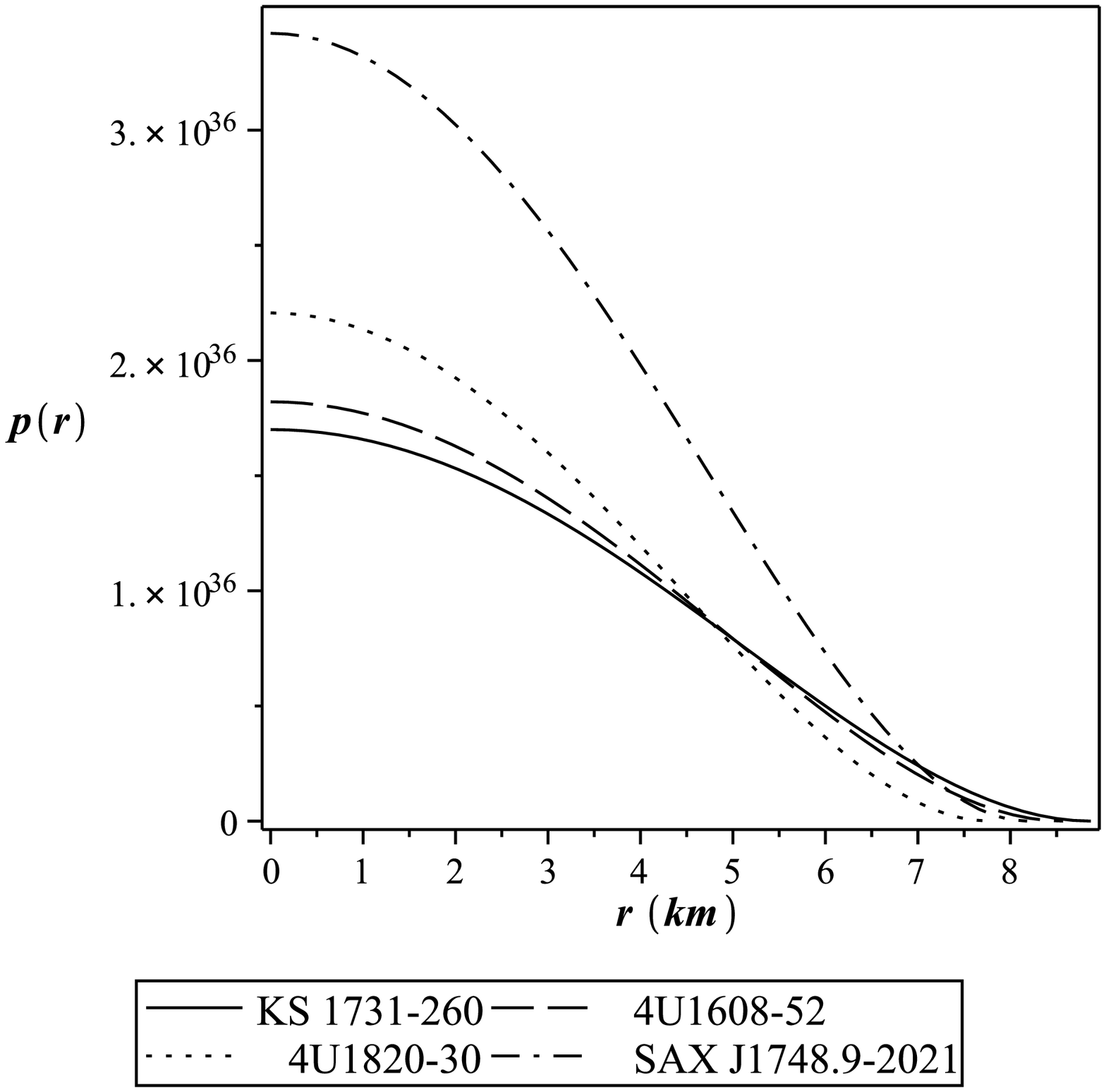}
\centering
\includegraphics[width= 6cm]{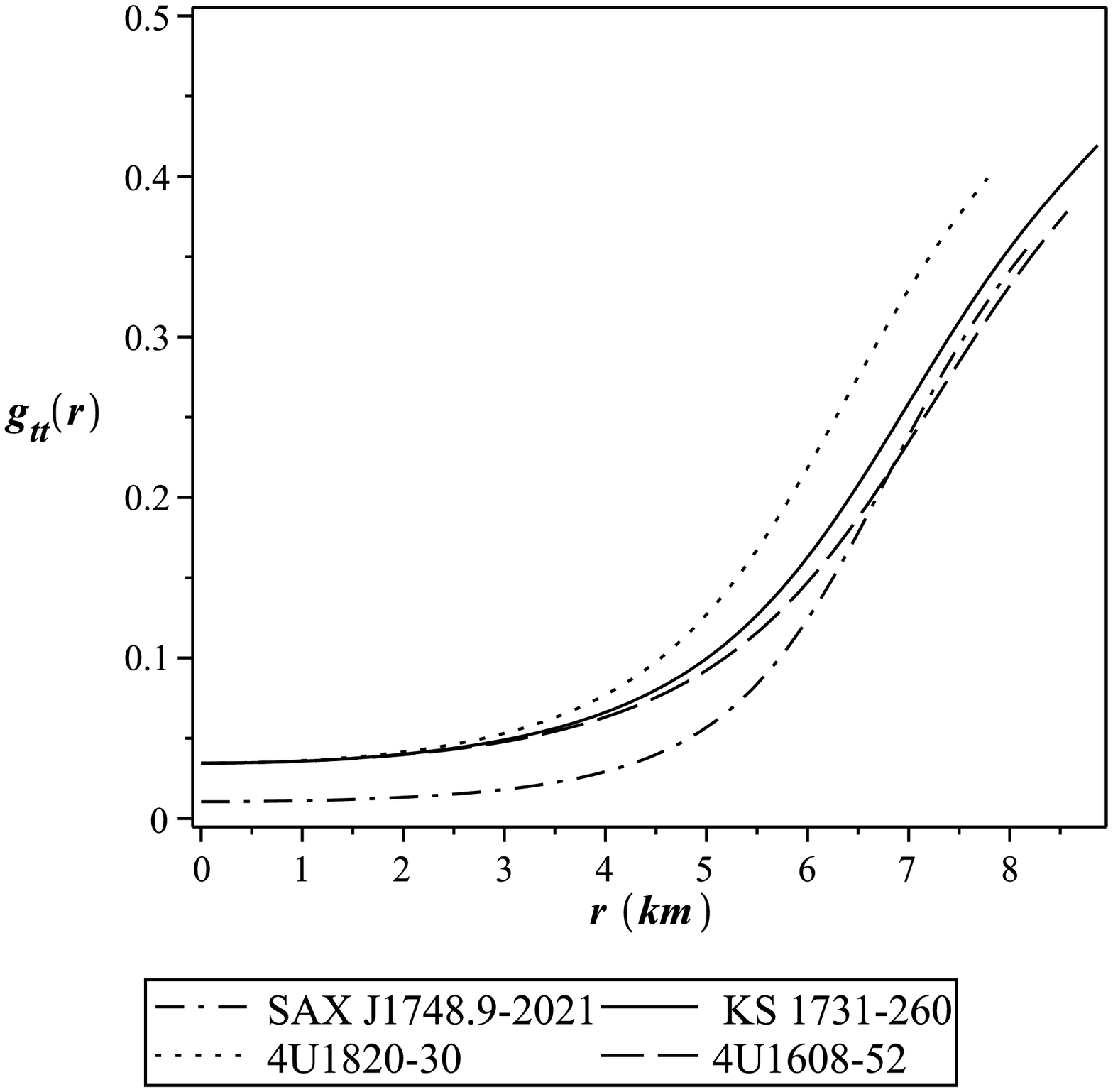}

\includegraphics[width= 6cm]{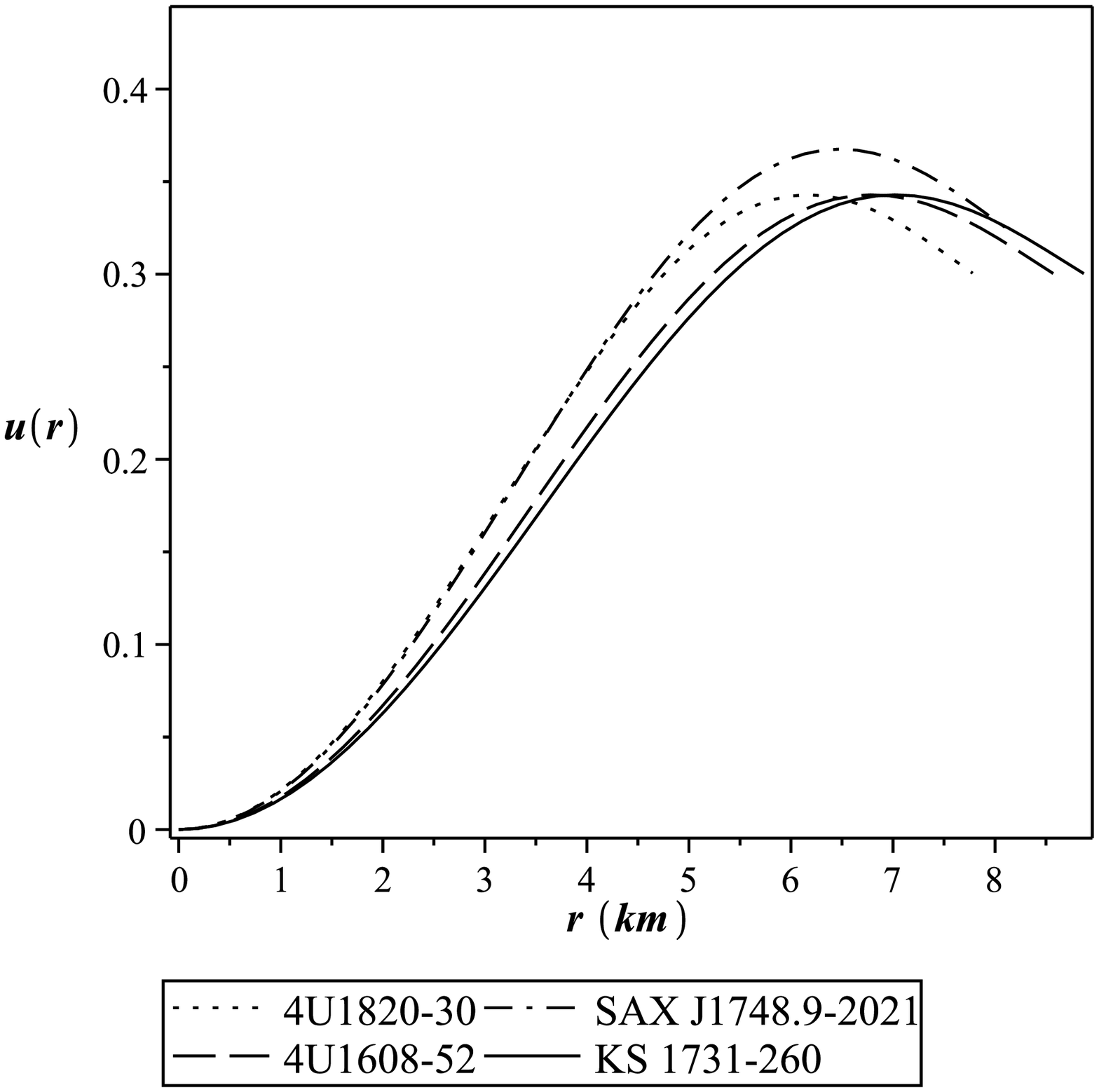}
\includegraphics[width= 6cm]{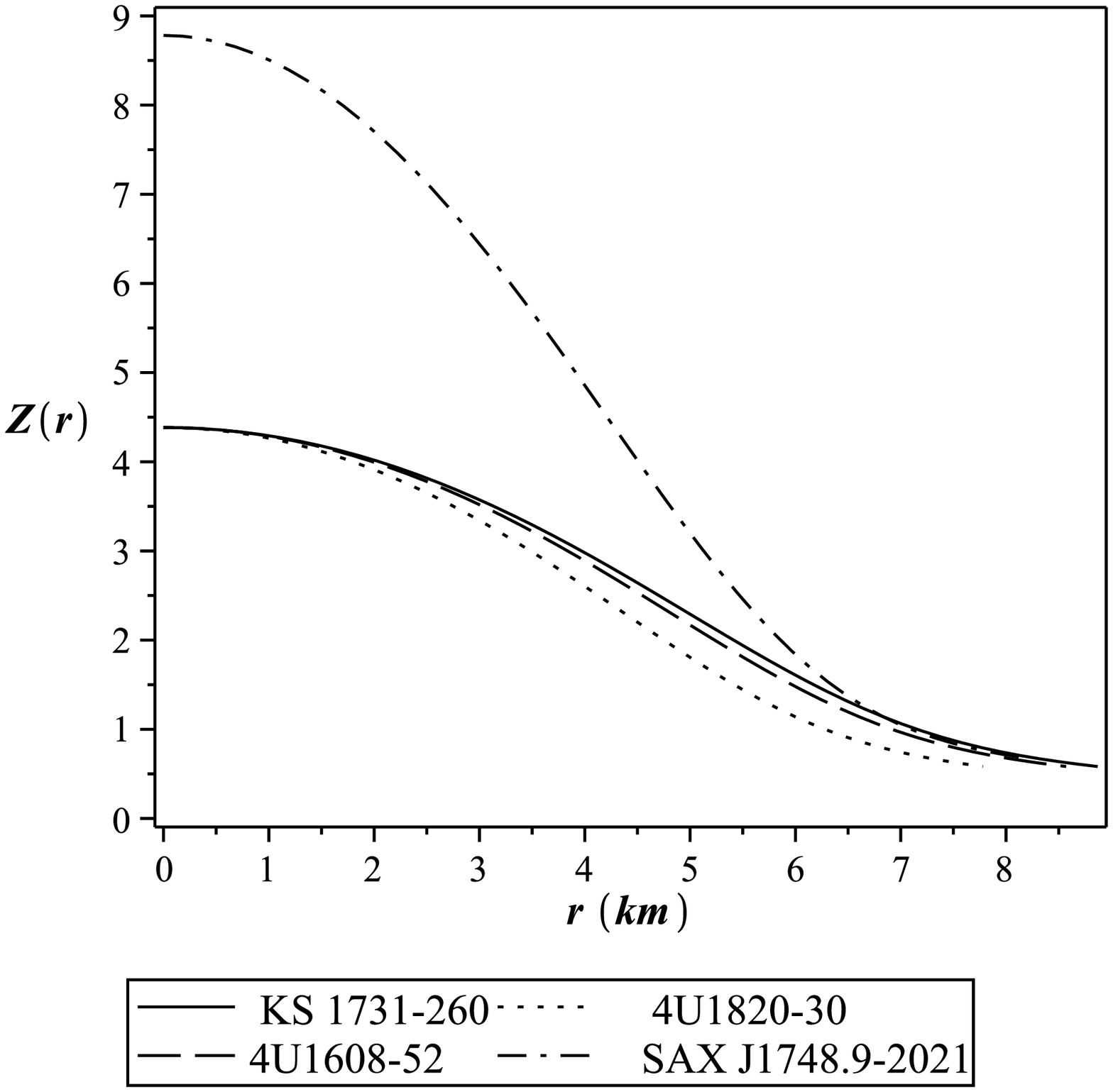}
\caption{ Solution III:  Density  in $gm/cm^3$ (upper left), pressure in $dyn/cm^2$ (upper right), the time-time component (middle), compactness factor (lower left) and redshift (lower right) as a function of the radial distance $r$.} \label{Fig5}
\end{figure}

The value of the physical parameters for the stars $RX~J185635-3754$~\cite{Pons2002}, $GS~1826-24$~\cite{Zamfir2012}, $4U/MXB~1728-34$~\cite{Majczyna2005} using Solution I are summarized in Table \ref{tbl-3} and for the stars $4U~1608-52$~\cite{Guver2010}, $4U~1820-30$~\cite{Guver2010a}, $KS~1731-260$~\cite{Ozel2012}, $SAX~J1748.9-2021$~\cite{Guver2013} using Solution III are summarized in Table \ref{tbl-4}. The variation of time-time component of the metric tensor, density, pressure, compactification factor and internal redshift with respect to the radial distance are presented by  Fig. \ref{Fig4} for Solution I and by Fig. \ref{Fig5} for Solution III. In Table \ref{tbl-2} we present the maximum possible mass of a star with the corresponding radius and surface redshift for different central densities.

\begin{table*}[h]
\centering \caption{Allowed ranges of the physical parameters for the presented model} \label{tbl-1}
\begin{tabular}{@{}lrrrrrrrrr@{}}
 \hline Soln. &$\Omega$ & $n$  & $M/R$ & $Z_{s}$  \\ \hline

I & $\Omega_{0}+\Omega_{1}$ & $8.4\leq n \leq 10.0$ & $0<\frac{M}{R}<0.319$ & $0<Z_{s}\leq0.67$ \\ 

II & $\Omega_{0}-\Omega_{1}$ & $8.4\leq n \leq 10.9$ & $0.320<\frac{M}{R}<0.442$ & $0.67 <Z_{s}\leq 1.95$ \\

III & $\Omega_{0}$ & Independent & $0.297<\frac{M}{R}<0.325$ & $0.57<Z_{s}\leq 0.69$ \\ \hline
\end{tabular}
\end{table*}

\begin{table*}[h]
\centering \caption{Set of values of the total mass and radius of stars having the maximum surface redshift for different central density} \label{tbl-2}
\begin{tabular}{@{}lrrrrrrrrr@{}} \hline                                                                                           
 Soln. & $\rho~$($g/cm^{3}$)  &$5\times 10^{14}$    & $1\times10^{15}$    & $5.5\times10^{15}$    & $Z_{s}^{Max}$  \\ \hline

I     & $[M(M_\odot), R(km)]$  & [6.50, 30.05] & [4.59, 21.25] & [1.96, 9.06]  &  0.67 \\ 

II    & $[M(M_\odot), R(km)]$  & [8.59, 28.71] & [6.07, 20.30] & [2.58, 8.66]  &  1.95 \\ 

III   & $[M(M_\odot), R(km)]$  & [6.66, 30.30] & [4.71, 21.42] & [2.01, 9.13] &  0.69 \\ \hline

\end{tabular}
\end{table*}

\begin{table*}
\centering \caption{Physical parameters of stars for Solution I} \label{tbl-3}
\resizebox{\columnwidth}{!}{
\begin{tabular}{@{}lrrrrrrrrr@{}}
\hline 
System               & $M$  & $n$ & $\omega$ & $R_i,R_o$ & $\rho_c\times10^{16}$ & $p_c\times10^{36}$ & $Z_c$ & $Z_s$ & $\frac{M}{R}$  \\ 
                     &\small $(M_\odot)$ &      &   & (km) & $(g/cm^{3})$ & $(dyn/cm^{2})$ &  &  &   \\
 \hline 
\small RX J185635-3754    & 0.90 & 8.4 & 0.51 & 6.00, 4.62  & $1.79$   & $8.26$ & 7.49 & 0.34 & 0.22 \\

\small GS 1826-24   & 1.70 & 8.5 & 0.47 & 12.85, 9.61 & $0.409$  & $1.73$ & 6.50 & 0.28 &  0.19 \\

\small 4U/MXB 1728-34        & 0.63 & 8.6 & 0.39 & 5.67,4.13  & $2.14$   & $7.53$ & 4.74 & 0.22 & 0.16 \\ 
             
\hline 
\end{tabular}} 
\end{table*}

\begin{table*}
\centering \caption{Physical parameters of stars for Solution III} \label{tbl-4}
\resizebox{\columnwidth}{!}{
\begin{tabular}{@{}lrrrrrrrrr@{}}
\hline 
System & $M$ & $\omega$ & $R$ & $\rho_c\times10^{15}$  & $p_c\times10^{36}$ & $Z_c$ & $Z_s$ & $\frac{M}{R}$ \\ 
 &\small~$(M_\odot)$ &  & (km) & $(g/cm^3)$  & $(dyn/cm^2)$ & &  & \\ 
\hline 
4U 1608-52       & 1.74 & 0.35 & 8.57 & $5.77$ & $1.82$ & 4.38 & 0.58 & 0.30 \\ 
 
4U 1820 30      & 1.58 & 0.35 & 7.78 & $6.99$ &  $2.21$ & 4.38 & 0.58 & 0.30 \\ 
 
KS 1731 260      & 1.80 & 0.35 & 8.87 & $5.39$ &  $1.70$ & 4.38 & 0.58 & 0.30 \\ 

SAX J1748.9-2021 & 1.78 & 0.559 & 8.18 & $6.79$ &  $3.42$ & 8.78 & 0.68 & 0.32 \\ 
\hline 
\end{tabular}}
\end{table*}

\begin{table*}
\centering \caption{Residual mass, binding coefficient and physical parameters corresponding to core-cust boundary of star} \label{tbl-5}
\resizebox{\columnwidth}{!}{
\begin{tabular}{@{}lrrrrrrrrr@{}}
\hline 
System & $Res[m]_{r=R}$ &  $\sigma$ & $r_{cc}~(km)$ & $\rho_{cc}~(g/cm^{3})$ & $p_{cc}~(dyn/cm^{2})$ \\ 
\hline
RX J185635-3754      & -2.284921      &  0.58713  &   3.09 & $7.30\times 10^{15}$ & $3.35\times 10^{36}$     \\ 

GS 1826-24           &  -0.395751    &  0.60995  &   6.55 & $1.62\times 10^{15}$ & $6.87\times 10^{35}$     \\ 

4U/MXB 1728-34       &  -0.391689    &  0.62061  &   2.86 & $8.32\times 10^{15}$ & $2.92\times 10^{36}$     \\
 
4U 1608-52           &  0.000355      &  0.46259  &   4.94  & $2.57\times 10^{15}$ & $8.11\times 10^{35}$    \\ 
 
4U 1820 30           &  0.005400      &  0.46302  &   4.49  & $3.11\times 10^{15}$ & $9.82\times 10^{35}$    \\ 
 
KS 1731 260          &  -0.001812     &  0.46302  &   5.12  & $2.39\times 10^{15}$ & $7.34\times 10^{35}$    \\ 
 
SAX J1748.9-2021     &  -0.000007     &   0.53900 &   4.72  & $3.02\times 10^{15}$ & $1.52\times 10^{36}$    \\ 
\hline 
\end{tabular}}
\end{table*}

\begin{figure}[!htp]
\centering
\includegraphics[width=8cm]{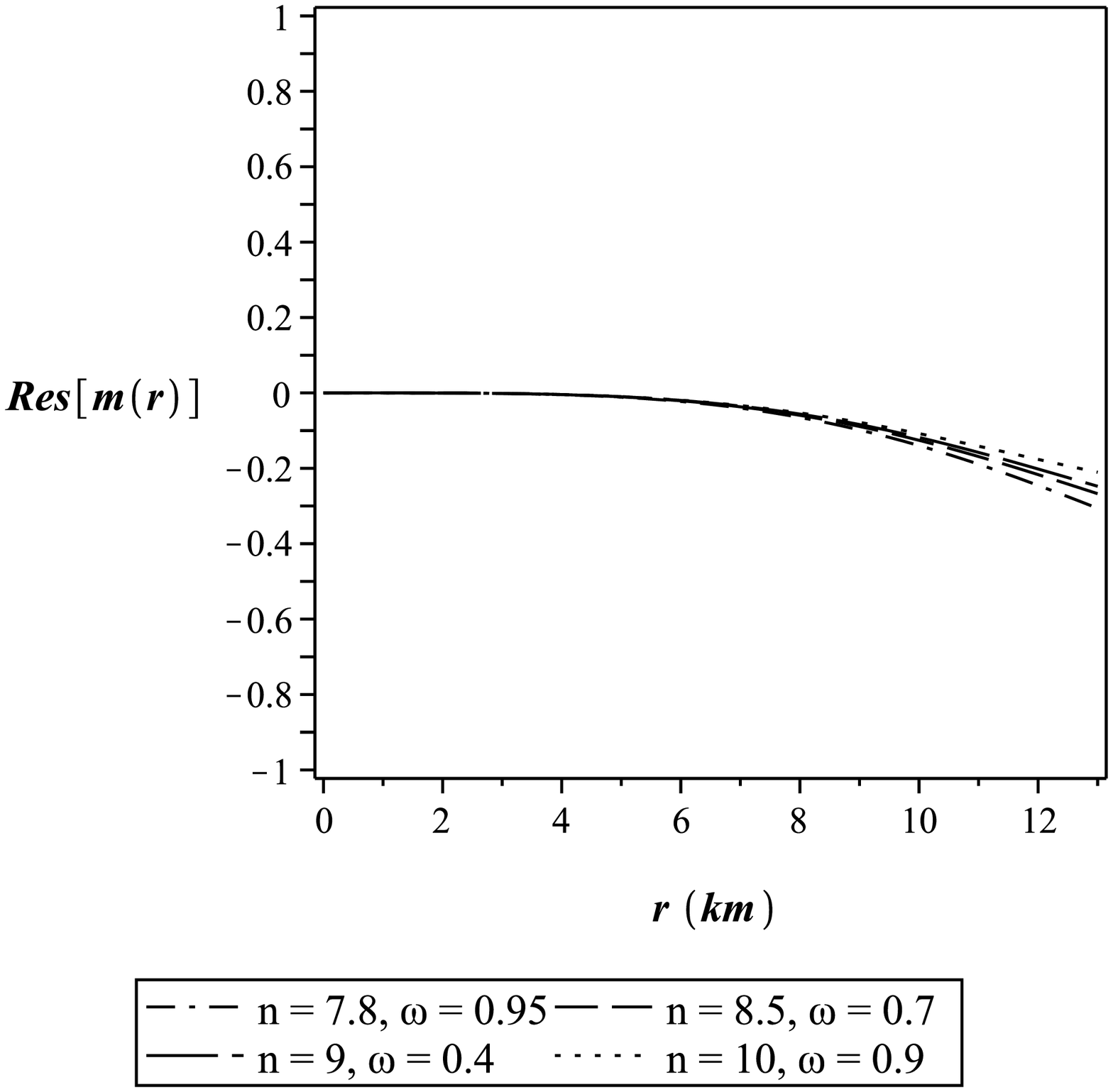} 
\caption{ Residual mass function  for $\rho_c=5\times10^{14}~g/cm^3$. }
	\label{Fig6}
\end{figure}

\section{Mass-Radius relation}
Different spectroscopic measurements of masses of neutron stars can be done mostly in radio observations and some in X-ray and gamma ray observations of pulsars in binary systems. There are also spectroscopic and timing measurements of radius of neutron stars. Based on the collections of data for mass and radius measurements of neutron stars it is seen that range of masses is from 1.17 $M_\odot$ to 2 $M_\odot$ and range of radii is from 10-11.5 km~\cite{Ozel2016}. Recent observation and interpretation of $GW~170817$ in different spectroscopic bands put constraint on both the upper and lower bound on tidal deformability parameter which consequently ruled out the extremely stiff and soft equation of states~\cite{Radice2018}. An analysis~\cite{Radice2018a} over the gravitational wave data and electromagnetic data of $GW~170817$ estimates a Bayesian parameter for the binary neutron star system subject to the assumption that the radii of neutron stars in $GW~170817$ are similar. In that analysis it is shown that considering only the gravitational wave data, neutron star of 1.4 $M_\odot$ cannot have radii smaller than 11.2 km. If electromagnetic data and systematic as well as statistical uncertainties are taken into account, the radius of  neutron star of 1.4 $M_\odot$ is expected to be $ 12.2^{+1.0}_{-0.8}\pm0.2$ km up to $90\%$ credible interval. In an another study~ \cite{Most2018} considering tidal deformability parameter of neutron stars in $GW~170817$, the constrained range for the radius of  purely hadronic neutron star of 1.4 $M_\odot$ is deduced to be  12-13.45 km. However, if phase transition is taken into account the range would be 8.53-13.74 km. Thus we see that there are various observational range of radii of neutron star depending on various type of measurements.

\begin{figure}[!htp]
\includegraphics[width=6cm]{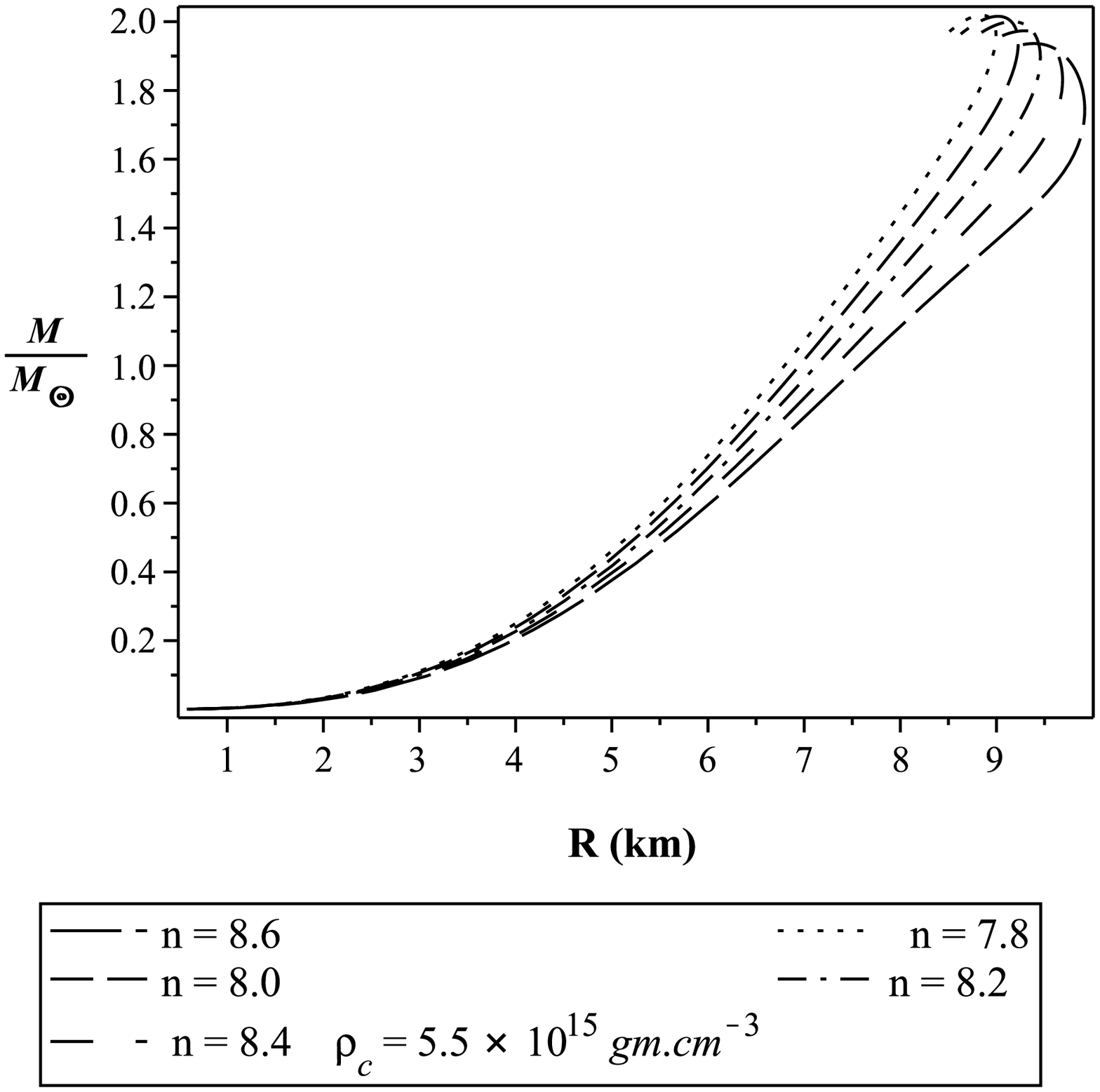} 
\includegraphics[width=6cm]{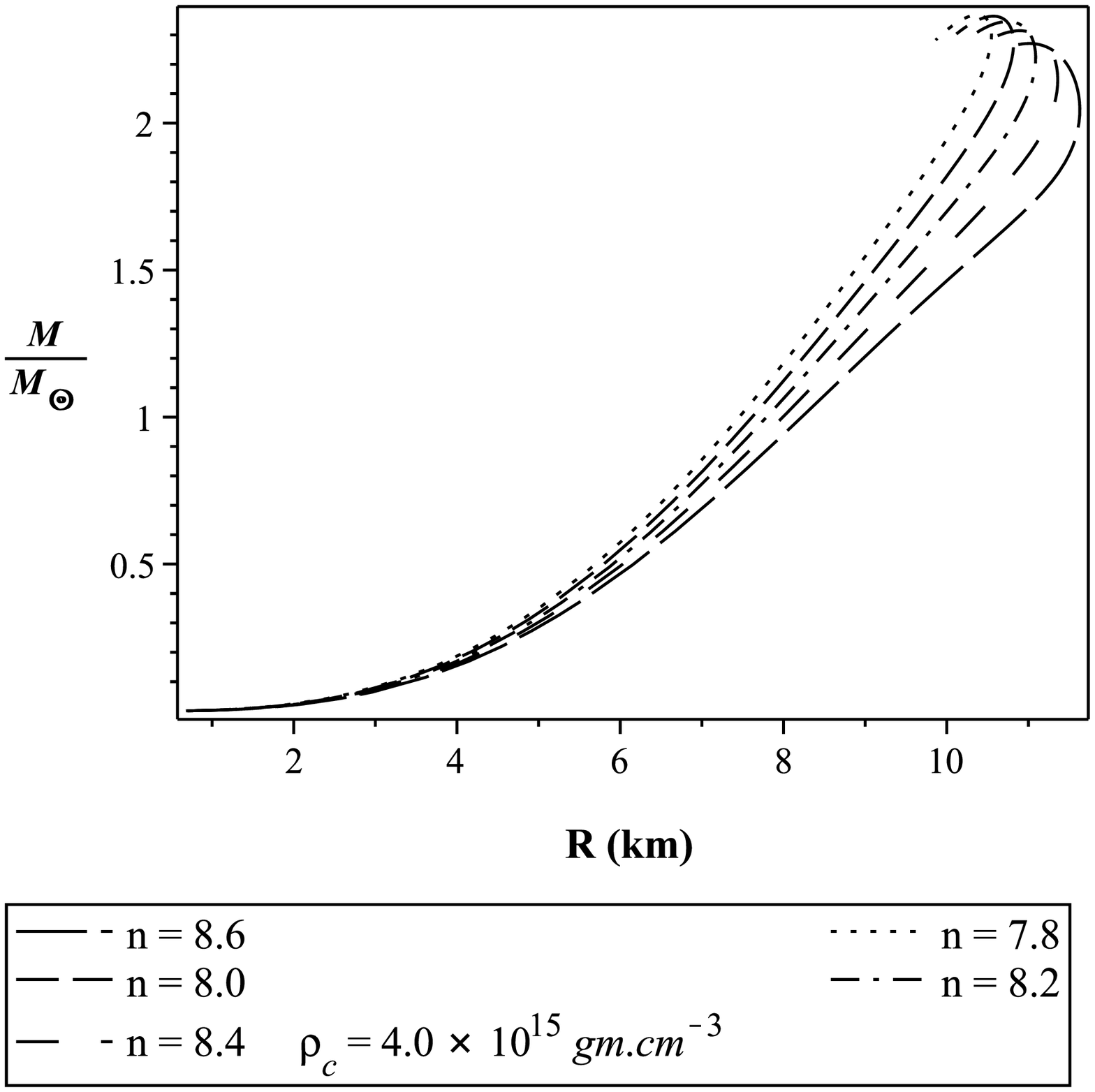}
 
\includegraphics[width=6cm]{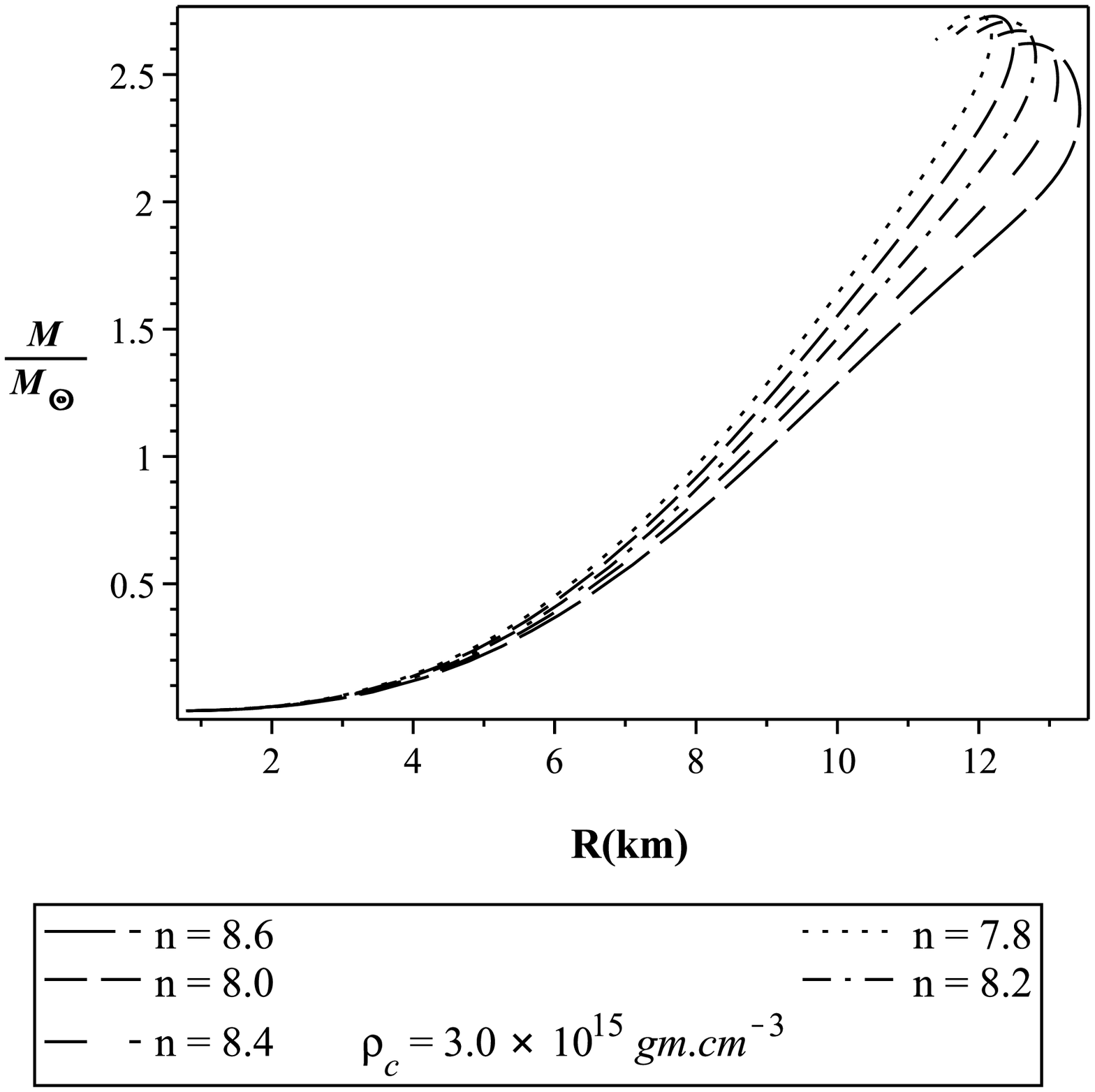} 
\includegraphics[width=6cm]{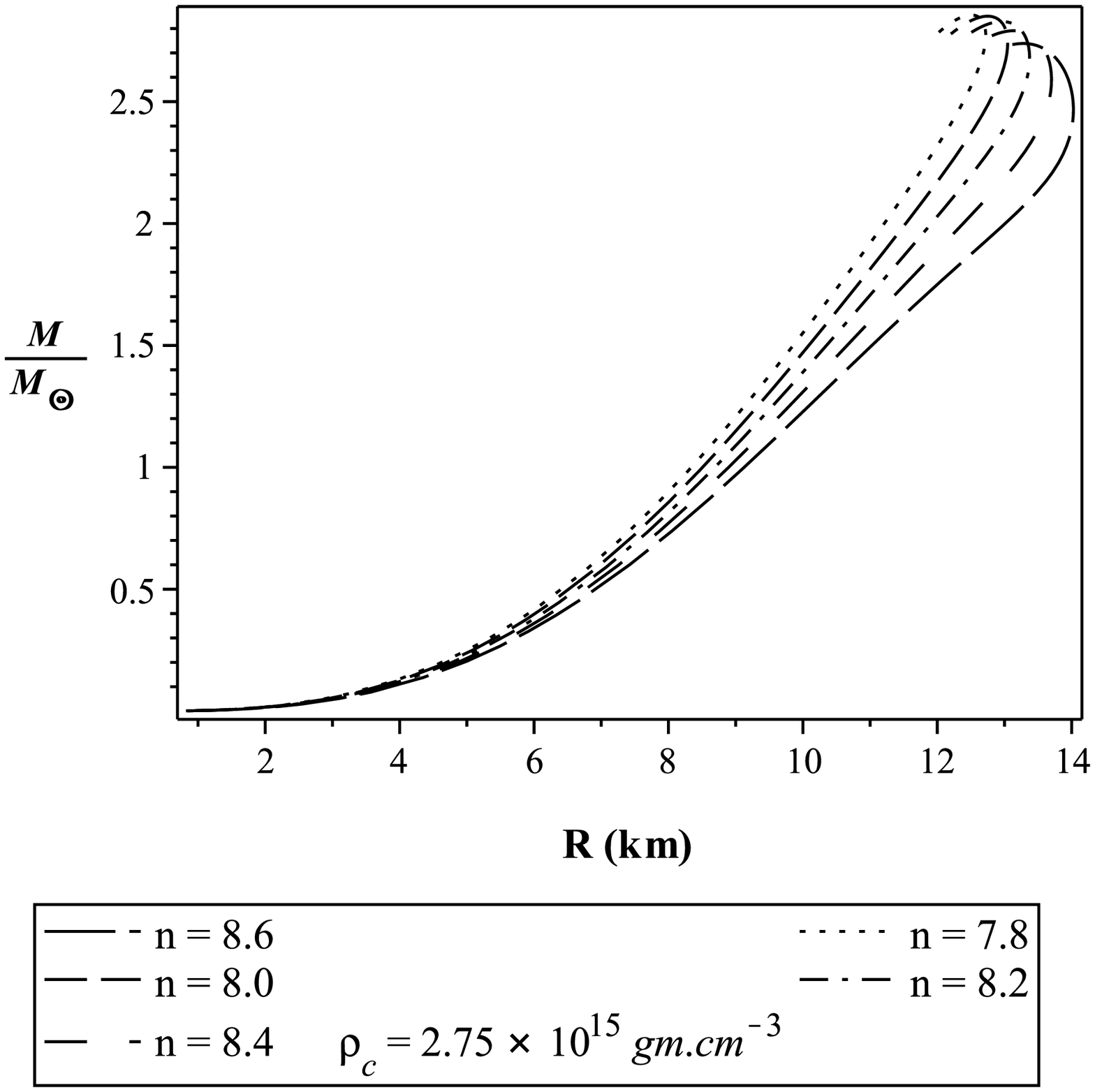} 
\caption{ Total mass (M) vs Radius (R) curves at different central densities such as  $5.5\times10^{15},~4.0\times10^{15},~3.0\times10^{15}, ~2.75\times10^{15}~ g/cm^{3}$ for one set $n=(7.8,~8.0,~8.2,~8.4,~8.6)$ using solution III. }
	\label{Fig7}
\end{figure}

\begin{figure}[!htp]
\includegraphics[width=6cm]{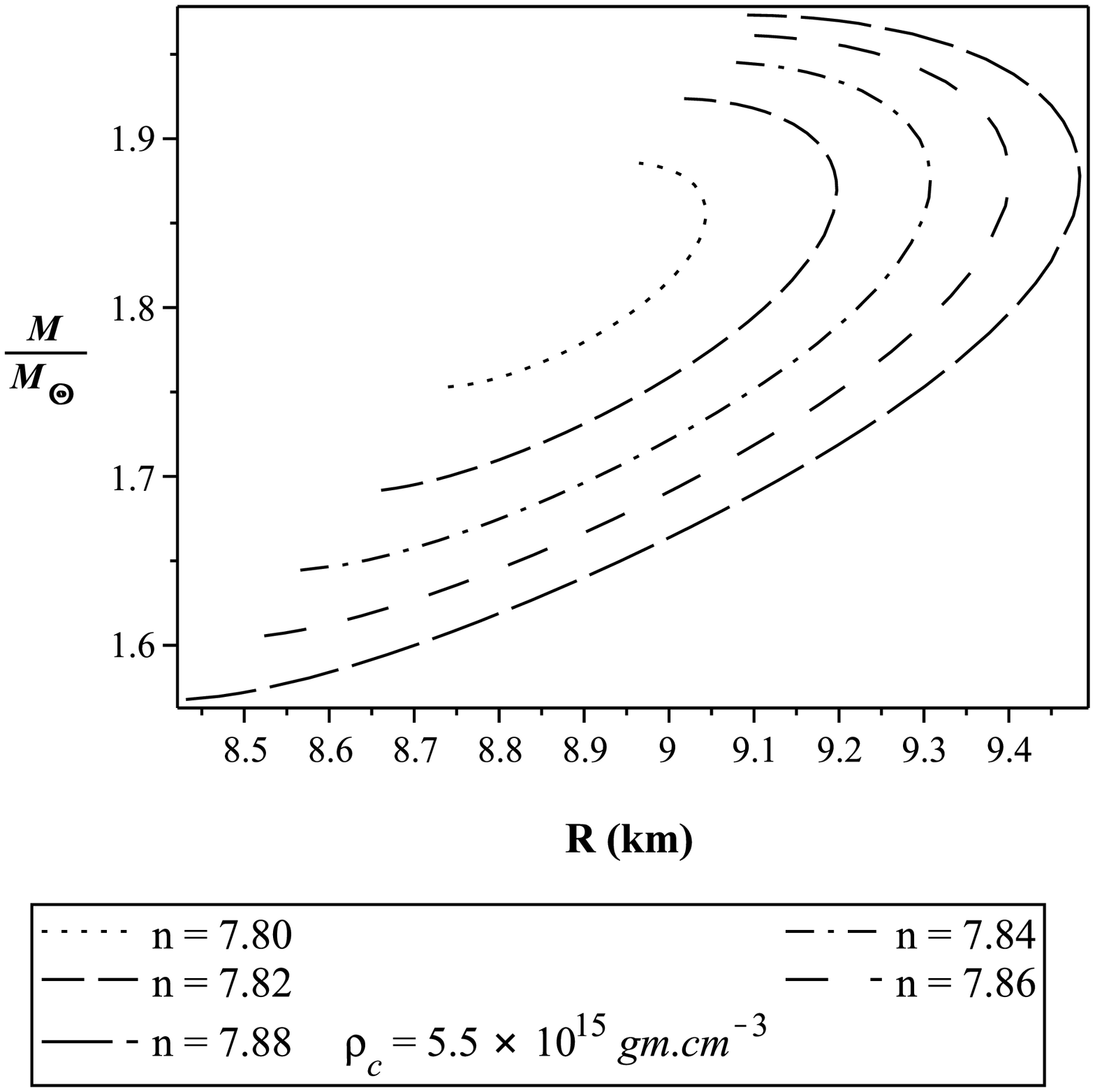} 
\includegraphics[width=6cm]{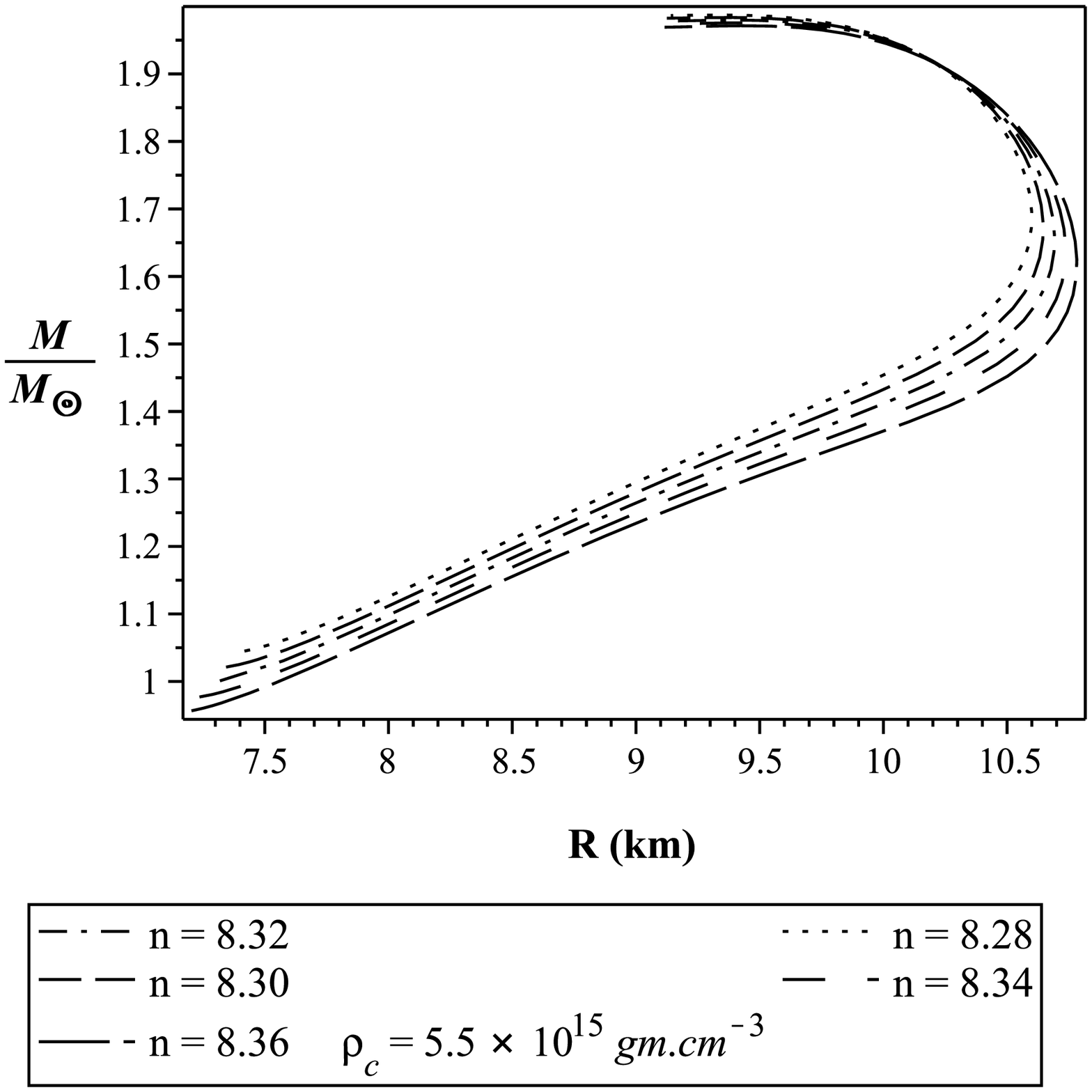} 

\includegraphics[width=6cm]{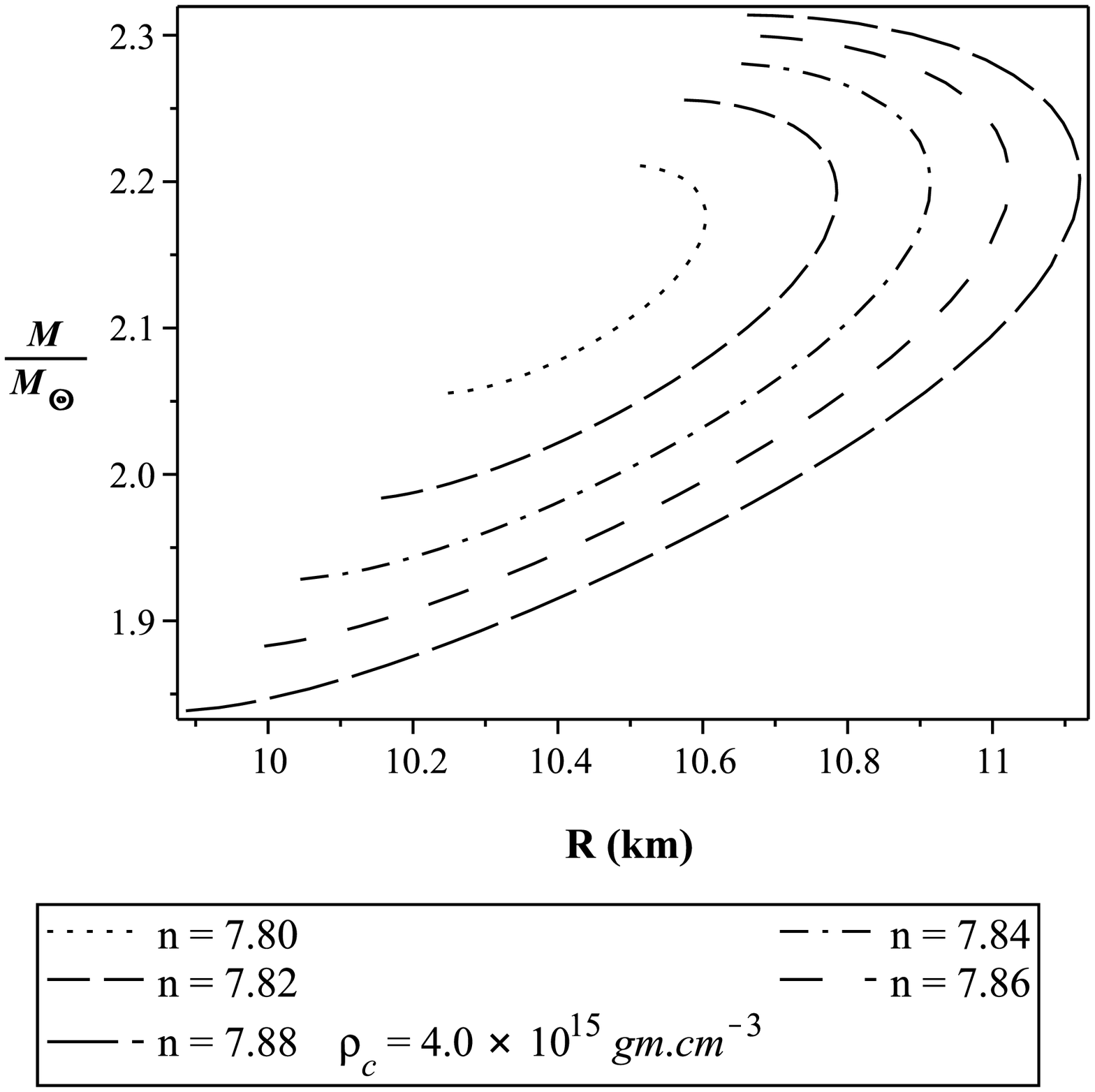} 
\includegraphics[width=6cm]{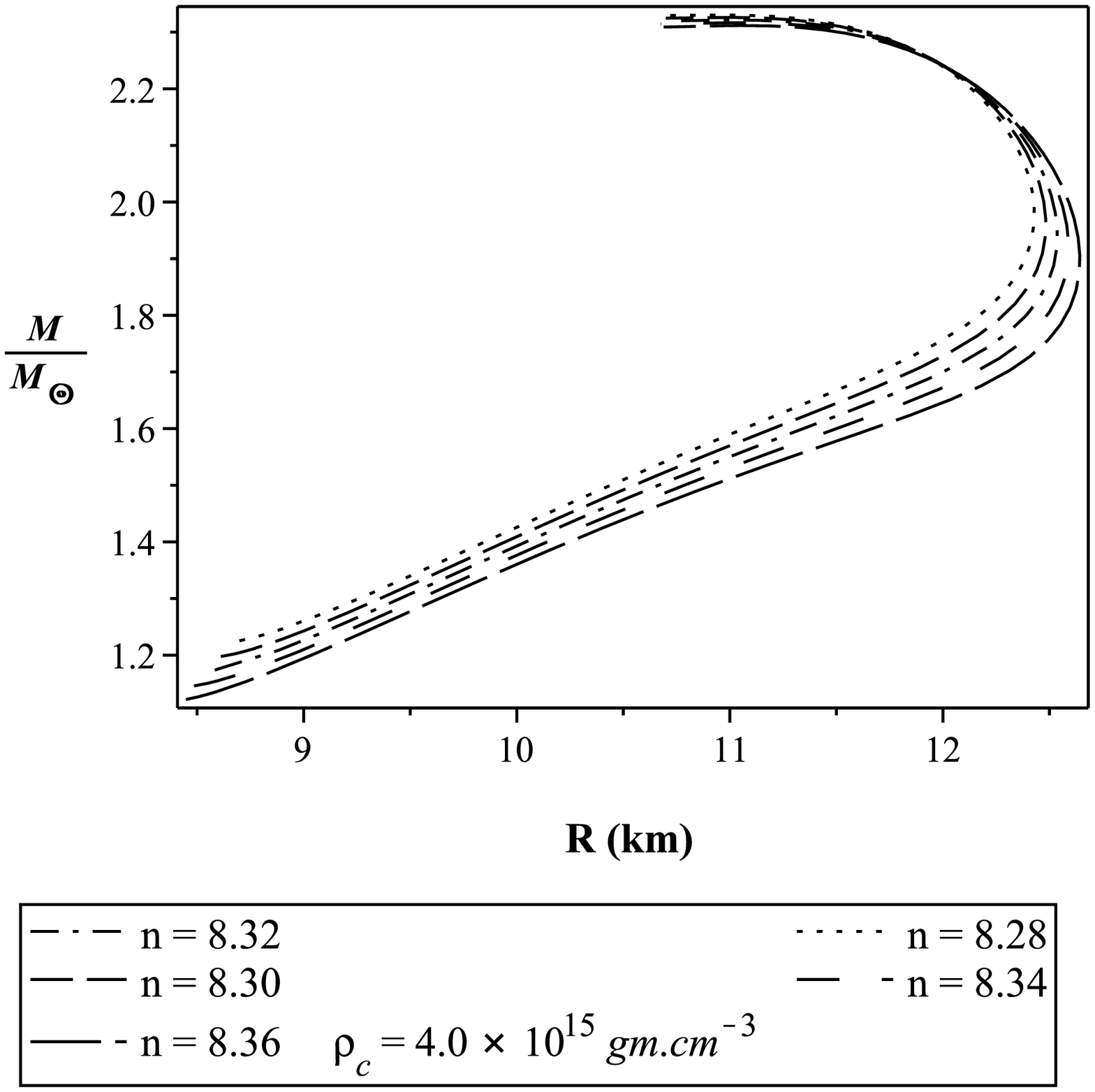}

\includegraphics[width=6cm]{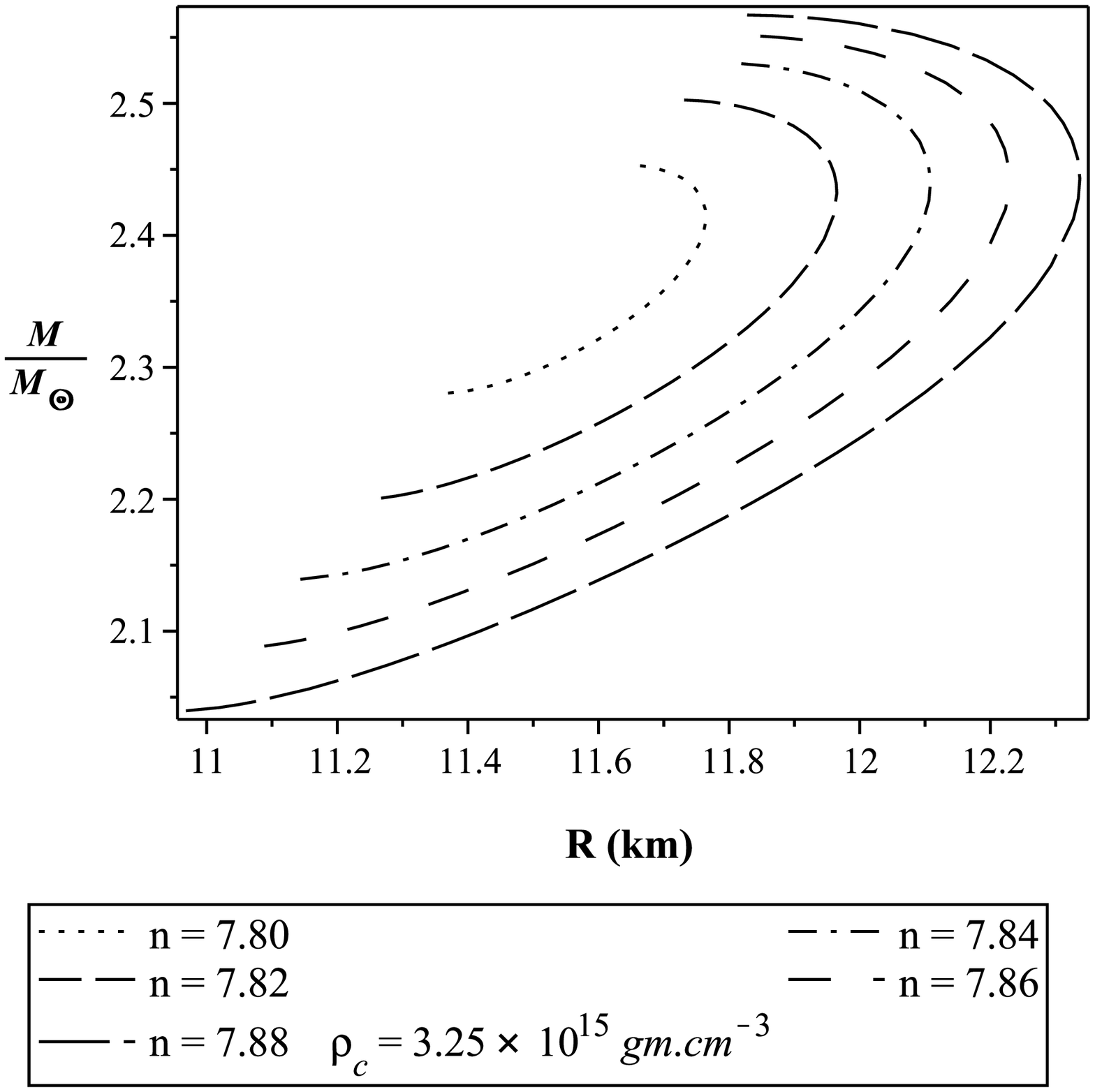} 
\includegraphics[width=6cm]{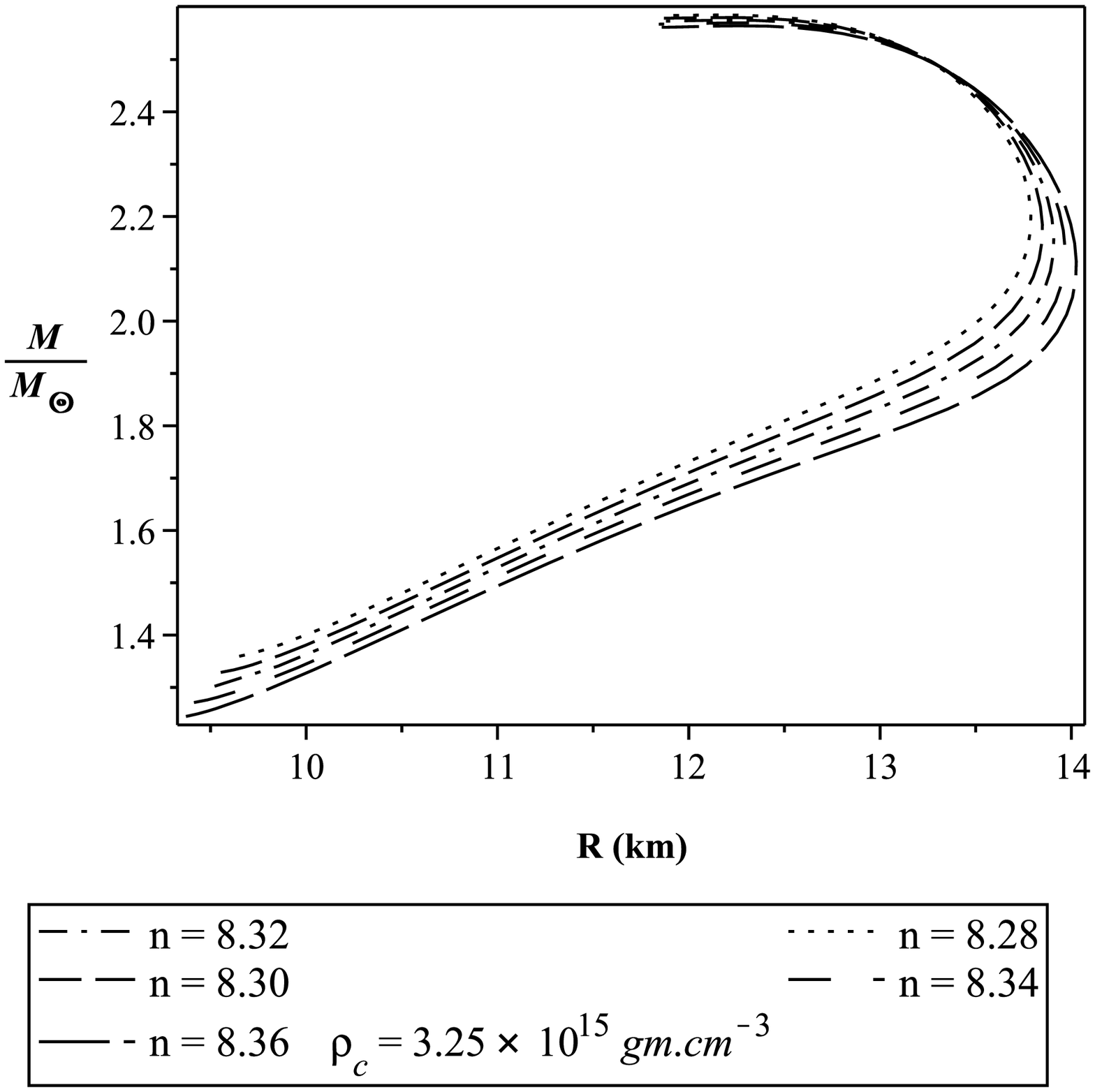}
\caption{  Total mass (M) vs Radius (R) curves at different central densities such as  $5.5\times10^{15},~4.0\times10^{15},~3.25\times10^{15}~ g/cm^{3}$ for two sets of $n=(7.80,~7.82,~7.84,~7.86,~7.88)$ and $(8.28,~8.30,~8.32,~8.34,~8.36)$ using solution II. }
	\label{Fig8}
\end{figure}

The presented stellar model is dependent on the choice of the model parameter $n$ and central density $\rho_c$. It is possible to obtain family of physical solutions for different values of the parameters ($n$, $\rho_c$) as the ranges of masses and radii have certain dependence on them and hence the observational ranges can be obtained just fine tuning the parametric values.  To show that our solutions for radii and masses indeed fall in the observational ranges, we have presented  variation of the total mass ($M$) with respect to the radius ($R$) of a star in Figs. \ref{Fig7} and \ref{Fig8} for different set of values ($n$, $\rho_c$) using solutions of the present relativistic model. We have also prepared a data Table \ref{tbl-6} showing range of masses and radii for different set of values of $n$ and $\rho_c$. From the mass-radius plots and the data Table \ref{tbl-6} it is revealed that for fixed $\rho_c$, the range of radii becomes wider as the value of $n$ increases whereas for fixed $n$, the value of maximum mass increases as $\rho_c$ decreases.

\begin{table*}[htbp!]
\caption{Ranges of bag constants for NS with quark core} \label{tbl-6}
\begin{tabular}{cccccccc}
\hline
 & \multicolumn{2}{c}{} & %
    \multicolumn{2}{c}{$\rho_c~$($g/cm^{3}$)} & \multicolumn{2}{c}{}\\
 & \multicolumn{2}{c}{$5.5\times 10^{15}$ } & \multicolumn{2}{c}{$4.0\times 10^{15}$ } & \multicolumn{2}{c}{$3.0\times 10^{15}$ }  \\
\cline{1-7}
 & $M~(M_\odot$) & $R~(km)$ & $M~(M_\odot$)  & $R~(km)$ & $M~(M_\odot$)  &$R~(km)$\\
$n$ & Range &Range & Range &Range &  Range &Range\\
\hline
7.8	& 1.75-1.88 		   & 8.74-9.05  & 2.05-2.21		   & 10.25-10.6 & 2.37-2.55		   & 11.84-12.25\\ 

 7.9  &  1.55-1.98	      & 8.50-9.46  &  1.80-2.30      & 9.85-11.21 &  2.08-2.68     &  11.38-12.93\\ 
    
 8.0  & 1.39-2.00 	   & 8.12-9.86  & 1.63-2.35	   & 9.54-11.58 & 1.88-2.71	   & 11.03-13.37 \\ 
   
 8.1  & 1.25-2.00 	   & 7.85-10.15 & 1.47-2.35 	   & 9.20-11.90  & 1.70-2.70 	   &  10.64-13.74\\ 
    
 8.2  & 1.14-1.99 	   & 7.62-10.42  & 1.33-2.34 	   & 8.93-12.18 & 1.54-2.69 	  &  10.32-14.07 \\  
     
 8.3 & 1.00-1.98 	   & 7.34-10.63 & 1.20-2.32	   & 8.62-12.46 & 1.38-2.68	   & 9.98-14.40\\ 
      
 8.4 & 0.91-1.96 	   & 7.02-10.85 & 1.07-2.29 	   & 8.25-12.73  & 1.24-2.65	   & 9.54-14.66\\ 
       
 8.5  & 0.80-1.93 	   & 6.79-11.06   & 0.95-2.26	   & 8.00-13.00 & 1.09-2.62	   & 9.24-15.00 \\  \hline
\end{tabular}
\end{table*}

In addition residual of the mass function or $Res[m]$ with respect to the radial distance  for some representative  values of the parameters ($n$, $\omega$)  are plotted to present the accuracy of approximate analytic solution of the TOV equation in Fig. \ref{Fig6}.  The small value of $Res[m]$ implies that accuracy is high enough to consider the approximate analytic solution to be physical one. In Fig. \ref{Fig6}, $Res[m]$ is zero almost up to 6 km and thereafter it has very small value compared to the mass function value. $Res[m]$ is also calculated for the observed stars and listed in Table 5. It is obvious that the accuracy of the solution can be made as high as possible with proper choice of the parameters $n$, $\omega$ and $\rho_c$.

\section{Conclusion}
In the present study the TOV equation is solved for the isotropic perfect fluid in the spherically symmetric spacetime  using homotopy perturbation method. We get approximate analytic solution for mass function. The mass function is governed mainly by the EOS parameter and model parameters. For any physically viable stellar model the parameters cannot have any value. We therefore determine the allowed ranges for parameters shown in Table \ref{tbl-1} following causality and stability condition for the star model. Three solutions are studied qualitatively and  the maximum mass, radius, maximum surface redshift, etc of a neutron star are predicted for different central density as shown in Table \ref{tbl-2}. For each solution to the model with a neutron star of known mass we find residual mass to check the accuracy of the solution and find binding coefficient to check the bound structure of the star. The residual mass, binding coefficient and the physical parameters corresponding to the core-crust boundary are given in Table \ref{tbl-5}. 

We provide here some important results and features of our model as follows:

(i) There is no special ansatz or form of metric potentials is assumed in the model. Only the value of $\omega$ and $n$ give the complete solution. The ratio of the total mass to the radius and surface redshift of a star only depends on $\omega$ and $n$. Also one can note that the total mass and radius are inversely proportional to the square root of the central density~\cite{Tolman1939,Nariai1950,Buchdahl1967,Lake2003}. The density, pressure, time-time component of the metric and other physical quantities are finite at the centre. So the solutions presented here are singularity free.

(ii) Using Solution I, we can describe the core of the NS which is physically valid. However, there is a shell of ($R_{o}-R_{i}$) in the crust region of NS which is non-physical as it contains negative density and thus violation of the energy conditions do occur. The residual mass being high the resulting solution I is not accurate enough to accept as a physical solution. On the other hand, the Solution II gives a physical solution for a star as density is positive throughout the region of the star and all the energy conditions are satisfied in this case.  This solution is very interesting as it predicts a NS to be highly compact and of high surface redshift. Similarly, the Solution III is completely physical and accuracy is very high as the residual mass is very small. However, the solution here only depends on the EOS parameter $\omega$. 

(iii) If we take Solutions II and III into consideration, our prediction is that the total mass to radius ratio of a NS should be constrained in the range $0.297<M/R<0.442$~\cite{Buchdahl1959} and surface redshift should lie in the range $0.57<Z_s<1.95$~\cite{Ozel2013} for the parametric values $8.4<n<10.9$ and $1/3<\omega<1$. The total mass as function of $\omega$ has a maxima at $\omega\approx0.6-0.75$. The maximum mass and radius for NS is predicted to be 2.01 $M_\odot$ and 9.13 km for the central density of $5.5\times10^{15}~g/cm^{3}$, $\omega=0.73$ and $Z_s=0.69$.

(iv) The mass-radius plots in Figs. \ref{Fig7} and \ref{Fig8} show the range of masses and radii of neutron stars within the observational range. Also Fig. \ref{Fig6} shows that approximate solution can be made accurate enough so that it could provide physical solutions for stellar model.

As a final comment: we note that though the mass function has approximate analytic solution but it gives non-singular and stable stellar configuration which can describe and study the properties of compact star (especially NS). \\

\section*{Acknowledgments} SR and FR are grateful to the Inter-University Centre for Astronomy and Astrophysics (IUCAA), Pune, India for providing Visiting Associateship under which a part of this work was carried out. FR is grateful to DST-SERB (EMR/2016/000193), Govt. of India for providing financial support. A part of this work was completed while AA was visiting IUCAA and the author gratefully acknowledges the warm hospitality and facilities at the library there.

\end{document}